\documentclass[journal]{IEEEtran}
\usepackage{amsmath,amsfonts}
\usepackage[colorlinks=true,linkcolor=blue,citecolor=blue,urlcolor=blue]{hyperref}
\usepackage{algorithmic}
\usepackage{algorithm}
\usepackage{array}
\usepackage[caption=false,font=normalsize,labelfont=sf,textfont=sf]{subfig}
\usepackage{textcomp}
\usepackage{stfloats}
\usepackage{url}
\usepackage{verbatim}
\usepackage{graphicx}
\usepackage{multirow} 
\usepackage{adjustbox}
\usepackage[dvipsnames]{xcolor}
\usepackage{pdflscape}
\usepackage{longtable}
\usepackage{booktabs}
\usepackage{ragged2e} 
\usepackage{amssymb}
\usepackage{comment}
\usepackage{float}
\usepackage{subcaption}
\usepackage{threeparttable}
\usepackage{tabularx}
\renewcommand\tabularxcolumn[1]{>{\raggedright\arraybackslash}p{#1}}
\usepackage{xr-hyper}           
\begin{document}

\title{Exploring human-SAV interaction using LLMs: The impact of psychological factors on user experience}

\author{
    \IEEEauthorblockN{Lirui Guo\textsuperscript{1}, Michael G. Burke\textsuperscript{2}, and Wynita M. Griggs\textsuperscript{1,2}}
    \thanks{*This work was supported by an Australian Government Research Training Program (RTP) Scholarship.}
    \thanks{$^{1}$Department of Civil and Environmental Engineering, Monash University, Clayton, VIC 3800, Australia.}
    \thanks{$^{2}$Department of Electrical and Computer Systems Engineering, Monash University, Clayton, VIC 3800, Australia.}
    \thanks{Email: \{Lirui.Guo, Michael.G.Burke, Wynita.Griggs\}@monash.edu}
}

\maketitle

\begin{abstract}
There has been extensive prior work exploring how psychological factors such as anthropomorphism affect the adoption of Shared Autonomous Vehicles (SAVs). However, limited research has been conducted on how prompt strategies in large language models (LLM)-powered conversational SAV agents affect users' perceptions, experiences, and intentions to adopt such technology. In this work, we investigate how conversational SAV agents powered by LLMs drive these psychological factors, such as psychological ownership, the sense of possession a user may come to feel towards an entity or object they may not legally own. We designed four SAV agents with varying levels of anthropomorphic characteristics and psychological ownership triggers. Quantitative measures of psychological ownership, anthropomorphism, quality of service, disclosure tendency, sentiment of SAV responses, and overall acceptance were collected after participants interacted with each SAV. Qualitative feedback was also gathered regarding the experience of psychological ownership during the interactions. The results indicate that an SAV designed to be more anthropomorphic and to induce psychological ownership improved users' perceptions of the SAV's human-like qualities, and its responses were perceived as more positive but also more subjective compared to the control conditions. Qualitative findings support established routes to psychological ownership in the SAV context and suggest that the conversational agent’s perceived performance may also influence psychological ownership. Both quantitative and qualitative outcomes highlight the importance of personalization in designing effective SAV interactions. These findings provide practical guidance for designing conversational SAV agents that enhance user experience and adoption.
\end{abstract}


\section{Introduction}\label{sec: intro}

Over the past few years, extensive research has been conducted to overcome technical, regulatory, and human-centric barriers to the adoption of Shared Autonomous Vehicles (SAVs), bringing them closer to practical reality. Combining shared mobility with automated driving technologies, SAVs are driverless taxi fleets composed entirely of fully autonomous vehicles \cite{daiImpactsIntroductionAutonomous2021}. SAVs offer numerous benefits, such as reduced car ownership and traffic congestion, lower environmental impact, decreased travel time costs, and enhanced accessibility for users \cite{wuDeepSuperficialAnthropomorphism2023, hudaUnderstandingValueAutonomous2023, pimentaExploringGapsResidential}. Practical advancements in SAV technology are already evident: Waymo provides autonomous ride-hailing services to the public in two U.S. cities \cite{WAYMOONEFuture}, while Apollo Go, a Chinese self-driving travel service platform, operates in 11 cities nationwide in China \cite{robotgoRobotgoAutonomousDriving2024}. Tesla has also announced plans for a Robotaxi, aiming to commence production before 2027 \cite{teslaWeRobot2024, TeslaSharesSink2024}. These examples demonstrate the practical advancements and transformative potential of SAV technology.

However, technological advancements alone are insufficient without public willingness to adopt SAVs \cite{zhangRolesInitialTrust2019}. A positive user experience (UX) is essential for the widespread adoption of SAV technology \cite{flohrChatTapComparing2021}. The high levels of autonomy associated with SAVs remove direct human control, which can lead to user unease if effective interaction mechanisms are not in place. This lack of control may negatively impact user comfort and overall acceptance of the technology \cite{flohrChatTapComparing2021, kangFeelingConnectedSmart2020, wangIdentifyingFactorsAffecting}. To bridge this gap, establishing effective interaction mechanisms is important. Current research tends to focus mainly on survey studies, driving simulations (e.g., driving simulators and virtual reality), and the physical appearance \cite{zhangRolesInitialTrust2019, zouRoadVirtualReality2021, xuAnalyzingScenarioCriticality2024, zhangToolsPeersImpacts2023, wangExploringImpactConditionally2024} of vehicles, while placing limited emphasis on the key aspect of conversational interactions between humans and SAVs, an area that becomes especially critical as autonomy advances. Such interactions, facilitated through natural language processing such as large language models (LLMs), can foster effective communication, user comfort, trust, and understanding in autonomous systems \cite{ruijtenEnhancingTrustAutonomous2018}. Notably, LLM-powered conversational interfaces may not only enhance user comfort by enabling natural dialogue but also endow systems with human-like traits, leveraging anthropomorphism to enhance perceived accuracy \cite{cohnBelievingAnthropomorphismExamining2024a}.

Anthropomorphism, the attribution of human-like characteristics to non-human entities \cite{sallesAnthropomorphismAI2020, wuDeepSuperficialAnthropomorphism2023}, is a key factor in human-SAV interaction that could affect user experience and intention to use SAVs. Research indicates that incorporating simplified human-like features, whether in appearance or behavior, can enhance users’ perceived service value, which in turn fosters loyalty and leads to a more positive interaction experience \cite{wuDeepSuperficialAnthropomorphism2023}. However, a greater level of anthropomorphism may decrease human trust and negatively affect human-SAV interaction quality, especially among respondents without vehicle ownership \cite{wuDeepSuperficialAnthropomorphism2023}. Therefore, finding the optimal level of anthropomorphism is essential for improving user acceptance.

As the name indicates, one of the key features of SAV is the sharing aspect, which involves the removal of private ownership. Over the past few decades, consumer behavior has shifted from the mantra ``we are what we own'' to ``we are what we can access'' \cite{merfeldCarsharingSharedAutonomous2019}. Psychological ownership (PO), the feeling of possession towards a target, whether legally owned, shared, or abstract, even in the absence of legal ownership  \cite{pierceStatePsychologicalOwnership2003, baxterPsychologicalOwnershipApproach2015}, is becoming increasingly significant for fostering engagement in access-based services. PO plays a vital role in how people interact with digital products and services, enhancing user engagement, satisfaction, and loyalty towards digital platforms \cite{kucinskasConsumerResponsesDiverse2024, liRolePsychologicalOwnership2020, liangPsychologicalOwnershipUsers2024, moonPlayerCommitmentMassively2013, daiImpactsIntroductionAutonomous2021}. Fostering a sense of PO when interacting with SAVs can make users feel a sense of personal connection to the shared vehicles. It could bridge the gap between private ownership and shared mobility services, facilitating the transition to shared mobility and ultimately increasing user acceptance in SAV systems \cite{leeAutonomousVehiclesCan2019, daiImpactsIntroductionAutonomous2021, orsot-dessiDeterminantsIntentionUse2023}.

In recent years, LLMs have become viable tools for enhancing user experience in various digital interaction domains. OpenAI’s ChatGPT, one of the most well-known LLMs, has gained significant popularity for its human-like conversational capabilities since its launch in November 2022 \cite{leiChatGPTConnectedAutonomous2023, duChatChatGPTIntelligent2023}. LLMs capable of facilitating natural interactions are increasingly influencing the field of intelligent vehicles. \cite{duChatChatGPTIntelligent2023} explored ChatGPT's potential in intelligent vehicle domains, such as driver assistance, autonomous driving, human-vehicle interactions, shared control, and enhancing the Driving Intelligence Quotient. Implementing LLMs in developing agents for communication with SAVs is expected to foster more natural dialogues between users and vehicles, ultimately enhancing user experience, confidence, safety, and comfort \cite{gaoChatChatGPTInteractive2023, duChatChatGPTIntelligent2023}. When integrated into SAVs, LLMs could play a significant role in triggering PO among SAV users and examining the effects of anthropomorphism. Applying strategies of PO and anthropomorphism when designing conversational agents for SAVs could optimize human-SAV interaction and enhance users’ perceptions and acceptance of SAVs. However, no prior studies have systematically examined how anthropomorphic and PO-driven conversational agents influence user experience and acceptance of SAVs. This paper investigates these effects by tailoring LLM-based conversational SAV agents to induce PO or anthropomorphism, aiming to identify strategies that enhance user adoption and satisfaction. 

This study aims to highlight differences in user experience across four types of SAV agents designed using different prompting strategies by evaluating factors such as quality of service, disclosure tendency, SAV response sentiment (i.e., polarity and subjectivity), and user acceptance. It demonstrates how strategic prompts can alter user perceptions, providing actionable insights for the development of conversational agents in SAVs that foster user experience and acceptance. By focusing on the integration of PO and anthropomorphism, this research fills a critical gap in the literature on human-SAV interactions and contributes to a broader understanding of how these elements affect technology adoption.


\section{Literature review}\label{sec:literature review}

\subsection{Psychological ownership}\label{Lit:PO}

Psychological ownership is the feeling that an object or entity is ``MINE'' \cite{pierceStatePsychologicalOwnership2003}. This feeling can occur for any objects that are visible, attractive, interesting, and experienced by an individual \cite{pierceStatePsychologicalOwnership2003, baxterPsychologicalOwnershipApproach2015}. This concept has gained prominence in consumer research as it can be experienced by all types of audiences even in the absence of physical or legal ownership \cite{thurridlHappyConsumptionPossessive2020}. With the rise of new technologies shifting the focus from tangible products to intangible experiences, psychological ownership has expanded beyond material possessions to encompass mobile technology services \cite{liangPsychologicalOwnershipUsers2024}.
In the context of car-sharing autonomous vehicles (AVs), where only one customer request (as an individual or a group) is being taken at a time \cite{narayananSharedAutonomousVehicle2020}, the customer temporarily takes ``ownership'' of an AV (i.e., while they are using this SAV service) without legally owning the vehicle. Thus, the concept of psychological ownership could potentially address a known SAV acceptance barrier caused by the absence of physical ownership \cite{leeAutonomousVehiclesCan2019}.
In this study, we define psychological ownership a user feeling that an SAV is “mine” for the duration of the ride \cite{liangPsychologicalOwnershipUsers2024}.

Psychological ownership has broad implications and leads to many benefits. In marketing research, psychological ownership is claimed to lead to greater happiness \cite{liRolePsychologicalOwnership2020}, increased purchase intention and willingness to pay \cite{kirkConsumerPsychologicalOwnership2018, hingstonWhatsMineMine2024}, more favorable attitudes toward domestic brands \cite{liangPsychologicalOwnershipUsers2024}, mediate the negative impact on consumer happiness \cite{zhaoSharedUnhappyDetrimental2023} and enhance stewardship behavior towards public goods \cite{peckCaringCommonsUsing2021}. In the context of digital technology, psychological ownership has been shown to improve engagement and satisfaction with digital products \cite{kucinskasConsumerResponsesDiverse2024}, foster loyalty to massively multiplayer online role-playing games \cite{moonPlayerCommitmentMassively2013}, enhance consumers’ valuation of the digital content \cite{kirkValueLurkingEffect2016}, and serve as a sustainable source of competitive advantage by actively involving customers in value co-creation processes \cite{kirkConsumerPsychologicalOwnership2018}. In the context of SAVs, psychological ownership is found to positively affect behavioral intention to use SAVs \cite{leeAutonomousVehiclesCan2019, daiImpactsIntroductionAutonomous2021, orsot-dessiDeterminantsIntentionUse2023}.

It is important to understand when and why users experience a sense of psychological ownership -- in other words, how psychological ownership can be triggered. Previous researchers claimed that the sense of ownership is rooted in four fundamental human motives (``roots''): (1) \textbf{effectance motivation} (the desire to interact effectively with the environment to achieve desired outcomes, fostering feelings of efficacy and pleasure), (2) \textbf{self-identity} (possessions help define oneself, express identity to others, and maintain self-continuity over time), (3) \textbf{home} (a sense of anchoring in time and space, offering familiarity, comfort, and security), and (4) \textbf{stimulation} (the need for arousal and activation) \cite{pierceTheoryPsychologicalOwnership2001, pierceHistoryPsychologicalOwnership2018}. These four motives can be seen as the reason for the existence of the psychological state of ownership rather than its cause \cite{pierceHistoryPsychologicalOwnership2018}. 

The development of psychological ownership occurs through three distinct ``routes'' (i.e., key experiences) -- \textbf{exercise of control}, such as physical control over an object \cite{atasoyDigitalGoodsAre2018}, control over the design process \cite{baxterPsychologicalOwnershipApproach2015, baxterOwnershipDesign2018}, or the perceived ability to influence a target’s behavior \cite{kirkDogsHaveMasters2019}; \textbf{intimate knowledge}, such as contamination concerns \cite{hingstonWhatsMineMine2024}, or repeated positive experiences in and with a place or target \cite{peckCaringCommonsUsing2021}; and, \textbf{investment of the self}, such as naming \cite{stonerNameGameHow2018}, designing \cite{kirkPropertyLinesMind2018}, customizing \cite{liRolePsychologicalOwnership2020}, and voting for \cite{fuchsPsychologicalEffectsEmpowerment2010} a product \cite{baxterPsychologicalOwnershipApproach2015, pierceHistoryPsychologicalOwnership2018, peckCaringCommonsUsing2021}. Beyond these routes, other strategies have been explored to trigger psychological ownership. This includes fostering positive affect during brand consumption \cite{thurridlHappyConsumptionPossessive2020}, using emotional narratives that incorporate elements of the three routes \cite{guoNewFrameworkPredict2025}, and employing actionable manipulations, such as placing signs like ``welcome to YOUR park” in public spaces \cite{peckCaringCommonsUsing2021}.

\subsection{Anthropomorphism}\label{Lit: A}

Anthropomorphism refers to the attribution of human-like traits, behaviors, or mental states to non-human entities, including objects, animals, brands, and technological artifacts \cite{sallesAnthropomorphismAI2020, wuDeepSuperficialAnthropomorphism2023}. In the context of SAVs, \cite{wuDeepSuperficialAnthropomorphism2023} classified anthropomorphism into two levels: \textbf{superficial anthropomorphism} and \textbf{deep anthropomorphism}. Superficial anthropomorphism represents lower-level dimensions such as physical appearances (e.g., friendly or aggressive), voice tones (e.g., friendly or cool), and vehicle movement patterns (elegant vs. rigid). In contrast, deep anthropomorphism encompasses higher-level dimensions like authenticity (lifelike or artificial), human-likeness (human-like or machine-like), naturalness (natural or synthetic), and awareness (conscious or unconscious).

Superficial anthropomorphism has been shown to bring numerous benefits to human-computer interactions. For instance, it fosters comfort, familiarity, and perceived trustworthiness during interactions, thereby enhancing the overall user experience with SAVs \cite{wuDeepSuperficialAnthropomorphism2023}. Additionally, anthropomorphic features can elicit more positive user responses by increasing a sense of connectedness when interacting with smart objects \cite{kangFeelingConnectedSmart2020}. It also positively influences users’ attitudes and facilitates the establishment of user–agent relationships \cite{wuDeepSuperficialAnthropomorphism2023}. On the other hand, deep anthropomorphism may reduce the quality of human-SAV interaction, particularly among low-income, low-education male respondents without vehicle ownership \cite{wuDeepSuperficialAnthropomorphism2023}. Conversely, some studies have shown that human-like qualities positively influence users’ perceived friendliness and safety, augmenting trust and overall acceptance of virtual assistant technology \cite{calahorra-candaoEffectAnthropomorphismVirtual2024}. These contrasting findings suggest that the impact of different levels of anthropomorphism may vary across contexts and user groups. Thus, it is important to explore how anthropomorphism can be effectively integrated into SAV interactions and its influence on user experience. 

\subsection{Conversational interfaces and large language models}\label{Lit: llm}

In high-level SAVs, specifically those at SAE Levels 4 and 5 where direct human control is largely or entirely removed, passengers primarily interact with the vehicles via digital User Interfaces (UIs), which can be conversational UIs, graphical UIs, or more directly via voice comments \cite{ruijtenEnhancingTrustAutonomous2018, flohrChatTapComparing2021, tekkesinogluAdvancingExplainableAutonomous2025}. Conversational UIs and command-based interfaces allow users to communicate with the vehicle using natural language, akin to conversing with a real driver \cite{mctearConversationalInterface2016}. However, if these interfaces fail to understand users’ requests or provide incorrect information, it can lead to negative interaction experiences and even serious issues \cite{folstadUsersExperiencesChatbots2020, sugisakiUsabilityGuidelinesEvaluation2020, flohrChatTapComparing2021}. Another key challenge is developing explainable autonomous vehicle systems to ensure effective human–vehicle communication. Such systems enhance user trust, promote transparency and accountability in the vehicle’s decision-making processes, and contribute to overall safety, user experience, and comfort \cite{tekkesinogluAdvancingExplainableAutonomous2025}.

Advancements in artificial intelligence (AI) have led to the development of conversational agents powered by LLMs, such as OpenAI’s ChatGPT. These models can generate human-like text, enabling more natural and engaging interactions \cite{flohrChatTapComparing2021, huaUseLargeLanguage2024}. One potential solution to enhance the effectiveness of the conversational ability of SAVs is the adoption of LLMs. Implementing conversational agents powered by LLMs with anthropomorphic personalities is expected to improve the system's understanding of user requests and increase user engagement. This integration can lead to an enhanced user experience and greater acceptance of SAVs. Additionally, inspired by growing interest in how different forms of engagement influence psychological ownership \cite{peckCaringCommonsUsing2021}, this study incorporates elements from the three established routes to psychological ownership into the design of conversational SAV agents to trigger users’ sense of ownership during interactions. 

Unlike traditional rule-based or menu-driven in-car chatbots, an LLM-powered SAV agent enables open-domain dialogue. This allows passengers to freely ask questions or make requests, such as requesting route changes, asking for local recommendations, or sharing emotions, without being confined to a predefined set of commands. Such flexibility fosters more natural, human-like interactions that adapt to each passenger's needs and context, better simulating the complexity and emotional nuances of real-world SAV use. 

Therefore, this research investigates how different anthropomorphic personalities and psychological ownership strategies, enabled by LLMs, influence human–SAV interactions, user experience, and acceptance.

\subsection{User experience factors}

In this project, the user experience (UX) is examined from the following aspects: user's disclosure tendency, user's perception of the SAV's service quality, user's perception of enjoyment when interacting with the SAV, the sentiment of the SAV responses, and user's acceptance of the SAVs. 

\subsubsection{Disclosure tendency}

Disclosure tendency (DT) refers to the willingness and behavior of individuals to reveal personal information to others, including computers \cite{zhangToolsPeersImpacts2023}. During human-computer interactions, sharing personal information can bring various benefits, such as personalized services, increased convenience, enhanced safety, the establishment of intimate relationships, smoother information exchange, and ease of communication. However, it can also pose risks, including privacy concerns, identity theft, financial fraud, and unauthorized secondary use of information \cite{songPredictorsConsumersWillingness2021, chengGoodBadUgly2022, zhangToolsPeersImpacts2023}. Interestingly, \cite{chengGoodBadUgly2022} found that in ride-sharing contexts, passengers tended to focus more on the benefits of disclosing personal information than on the associated risks. 

Moreover, \cite{zhangToolsPeersImpacts2023} observed that participants were more willing to share personal information with intelligent agents (IAs) designed to play a servant role when these agents exhibited human-like features (e.g., a friendly tone, relatable visual design) compared to IAs with robot-like characteristics (e.g., mechanical voice, robotic appearance). In the context of SAVs, users are required to disclose personal information such as their travel routes. To achieve a more personalized travel experience, users may also need to share additional preferences, such as their music tastes, to customize SAV settings. If they perceive the SAV as their own vehicle, they may be more willing to share their personal information. Accordingly, we examine how conversational agent design, specifically psychological ownership and anthropomorphic cues, influences users’ willingness to disclose personal information during interactions with SAVs

\subsubsection{Quality of service}

Quality of service (QoS) is a key factor in evaluating user experience with a product or service. Satisfied users are more likely to engage in positive word-of-mouth, exhibit a higher willingness to pay, and result in an increased intention to continue using the product or service \cite{daiImpactsIntroductionAutonomous2021, songPredictorsConsumersWillingness2021}. Research has shown that anthropomorphism can significantly influence the quality of consumer-AI interactions, with users tending to engage for longer periods when interacting with human-like chatbots or AI devices \cite{wangSmartphonesSocialActors2017}. Additionally, QoS has been identified as a key factor in users’ willingness to share personal information \cite{songPredictorsConsumersWillingness2021} and serves as a mediator between anthropomorphic characteristics and acceptance \cite{pelauWhatMakesAI2021}. 

\subsubsection{Perceived enjoyment}

Perceived Enjoyment (PE) refers to the user's feelings of pleasure and joy during interactions with SAVs \cite{songPredictorsConsumersWillingness2021}. Research on human-robot interactions has shown that enjoyable interactions can lead to a greater willingness to engage and share information \cite{songPredictorsConsumersWillingness2021}. Similarly, in the context of human-SAV interactions, it can be assumed that if users find the SAV enjoyable to interact with, their overall user experience is likely to improve.

\subsubsection{Sentiment of the SAV responses}

Sentiment analysis, a combination of Natural Language Processing, Computational Linguistics, and text analysis, focuses on extracting and understanding human sentiments from textual data \cite{bontaComprehensiveStudyLexicon2019}. It has been widely adopted across academia, businesses, governments, and other organizations \cite{wankhadeSurveySentimentAnalysis2022}. Sentiment analysis typically involves two classifications: subjectivity classification, which provides a \textbf{subjectivity} score indicating whether a text is subjective or objective, and sentiment classification, which delivers a \textbf{polarity} score reflecting whether a text is positive or negative \cite{wankhadeSurveySentimentAnalysis2022, wangSentimentClassificationContribution2014}. In the context of human-SAV interactions, these two measures (i.e., polarity and subjectivity) can be utilized to evaluate the sentiment of SAV responses.

\section{Method}\label{sec: method}

This section outlines the methodology employed to investigate the effects of psychological ownership and anthropomorphism in human-SAV interactions, contextualized within experimental settings in Victoria, Australia. The subsequent subsections detail the design of the SAV UI, LLM prompts, user study framework, hypotheses, and analytical approaches used to assess user experiences and acceptance.

\subsection{SAV User Interface design}

A conversational SAV UI prototype was designed to serve as the primary interface of the conversational SAV agents for investigating the effects of psychological ownership and anthropomorphism on user experience. The `gradio’ package was used to build the UI, and the `openai’ package was used to access the ``gpt-3.5-turbo'' model using the OpenAI API in Python \cite{gradio, openaiOpenAIPythonAPI2024}. The selection of the ``gpt-3.5-turbo'' model for this project was guided by its conversational abilities, cost-effectiveness, and high speed, making it the ideal choice for simulating human-SAV interactions \cite{openaiModels}. 


\subsection{SAV prompt design}

Four distinct types of SAV agents were designed, each utilizing different prompt strategies to explore the effects of psychological ownership and anthropomorphism on user experience and acceptance. The prompts were intentionally concise yet varied, integrating engagement and perception cues to simulate realistic, natural-language interactions. The aim of designing three distinct personas and allowing users to choose their preferred interaction style is to enhance user engagement and foster psychological ownership. Seven iterative pilot trials refined the wording to ensure that each agent’s communication style, functionality, and intended user experience were conveyed clearly. All prompts explicitly instructed the SAVs to strictly follow both Australian national laws and the specific regulations set forth by the State of Victoria, attempting to ensure safety and compliance at all times. The detailed descriptions and wording of the prompts can be found in the Supplemental Materials Section~\ref{S-Appendix: prompt list}. The inclusion of these variations allowed for a comparative analysis of how different prompts influence user engagement, satisfaction, and psychological ownership.

\subsubsection{Standard (i.e., the control group)}

The baseline prompt was designed to represent a standard SAE Level 5 SAV without emphasizing psychological ownership or anthropomorphic features. It included the core functionalities necessary for managing daily trips, such as adjusting climate preferences, controlling navigation, and handling media settings.

\subsubsection{Standard + [Psychological Ownership] prompt}

The second group incorporated strategies to foster a sense of psychological ownership by following the three PO routes introduced in Section \ref{Lit:PO} -- exercise of control, intimate knowledge, and investment of the self. The large language model was instructed to ensure users perceive the SAV as ``their own private autonomous vehicle'' by emphasizing control over vehicle operations, to share intimate knowledge about the SAV, and encourage users to invest themselves in its use. The goal was to enhance user attachment and engagement, and develop a strong sense of psychological ownership while users interacted with the vehicle.

\subsubsection{Standard + [Anthropomorphism] prompt}

This prompt employed three different anthropomorphic personalities to create a more human-like and personalized interaction style. The SAV was designed to embody one of three distinct personas: cool and sophisticated, friendly and engaging, or sassy and tired. Participants selected their preferred personality at the start of the user study (before all interactions).

\subsubsection{Standard + [Psychological Ownership] prompt + [Anthropomorphism] prompt}

The last group combined the strategies from the previous two groups, integrating PO routes with anthropomorphic personalities. The objective was to evaluate the combined effect of these approaches on user perception and user experience.

\subsection{User study design}

A user study was conducted with ethical approval from the Monash University Human Research Ethics Committee (MUHREC)\footnote{Project ID: 40485}. Participants were invited to interact with all four SAV groups (in random order). 
A list of fifteen sample user inputs (e.g., ``Adjust the seat, I want to lie down'') was provided, but participants were also informed that they were free to interact with the SAVs in any way they desired, including making requests or asking questions.
After each interaction, participants were asked to fill out a survey about their interaction experience covering the following aspects: their feelings of psychological ownership, the perceived anthropomorphism of the SAV, quality of service, their disclosure tendency, polarity (i.e., whether the responses of the SAVs were positive or negative), subjectivity (i.e., whether the responses of the SAVs were subjective or objective), enjoyment, and SAV acceptance. The questionnaire items used are summarized in Table \ref{tab:questionnaire}. Qualitative data was collected from open-ended questions about additional features or experiences with the vehicle that would contribute to psychological ownership, the benefits of sharing personal information, and interview responses about participants' perceived psychological ownership.
The user studies were held from April 30, 2024, to July 2, 2024. In total, 50 participants were invited, and 2,136 chats (a question-answer interaction) were collected. On average, each participant spent approximately one hour and 15 minutes completing the entire study. In this research project, the targeted population was selected to be young and highly educated, as this group may represent potential early adopters of technologies like SAVs due to their openness to innovation and higher levels of education \cite{guoNewFrameworkPredict2025}.

\begin{table*}
\centering
\caption{Questionnaire Items and Sources.}
\label{tab:questionnaire}
\resizebox{\textwidth}{!}{%
\begin{adjustbox}{max width=\textwidth}
\begin{tabular}{lp{4cm}}
\toprule
\textbf{Questionnaire items} & \textbf{Modified from} \\
\midrule

\multicolumn{2}{l}{\textbf{Psychological Ownership (PO)---Measured from 1 (Strongly Disagree) to 7 (Strongly Agree)}} \\
PO1: I feel like the vehicle is mine & \multirow{4}{*}{\cite{leeAutonomousVehiclesCan2019, liCanAIChatbots2023}} \\
PO2: I feel a very high degree of personal ownership of the vehicle & \\
PO3: Interacting with the SAV interface enhances my feeling of ownership over the vehicle & \\
PO4: I feel like I own the vehicle & \\
\addlinespace[2pt]

\multicolumn{2}{l}{\textbf{Anthropomorphism (A)---Measured from 1 (Strongly Disagree) to 7 (Strongly Agree)}} \\
A1: I believe that the SAV has its own decision-making power & \multirow{4}{*}{\cite{pelauWhatMakesAI2021, chiDevelopingFormativeScale2021}} \\
A2: I believe that the SAV has its own personality & \\
A3: I believe that the SAV has its own preferences and mood & \\
A4: I believe that the SAV has its own emotions & \\
A5: I believe that the SAV device can feel compassion & \\
\addlinespace[2pt]

\multicolumn{2}{l}{\textbf{Quality of Service (QoS)---Measured from 1 (Strongly Disagree) to 7 (Strongly Agree)}} \\
QoS1: I am satisfied with the SAV service. & \multirow{3}{*}{\cite{zhangToolsPeersImpacts2023}} \\
QoS2: I find communication with the SAV to be pleasant. & \\
QoS3: I would recommend the SAV to others. & \\
\addlinespace[2pt]

\multicolumn{2}{l}{\textbf{Disclosure Tendency (DT)---Measured from 1 (Strongly Disagree) to 7 (Strongly Agree)}} \\
DT1: I am willing to disclose my daily route with the SAV. & \multirow{3}{*}{\cite{zhangToolsPeersImpacts2023}} \\
DT2: I am willing to disclose my tastes in music with the SAV. & \\
DT3: I am willing to disclose my feelings, including negative emotions like anxiety, with the SAV. & \\
\addlinespace[2pt]

\multicolumn{2}{l}{\textbf{Polarity (Pol)---Measured from -1 (Negative) to 1 (Positive)}} \\
Pol: Overall, how do you rate the SAV responses, negative or positive? & By definition, \cite{wangSentimentClassificationContribution2014} \\
\addlinespace[2pt]

\multicolumn{2}{l}{\textbf{Subjectivity (Sub)---Measured from 0 (Objective) to 1 (Subjective)}} \\
Sub: Overall, do you think the SAV responses are objective or subjective? & By definition, \cite{wangSentimentClassificationContribution2014} \\
\addlinespace[2pt]

\multicolumn{2}{l}{\textbf{Perceived Enjoyment (PE)---Measured from 1 (Strongly Disagree) to 7 (Strongly Agree)}} \\
PE1: How fun do you feel your interaction with this SAV was? & \multirow{2}{*}{\cite{pelauWhatMakesAI2021}} \\
PE2: Do you think ``fun'' is a factor that would influence your intention to use SAVs? & \\
\addlinespace[2pt]

\multicolumn{2}{l}{\textbf{Behavioral Intention to Use (BI)---Measured from 1 (Strongly Disagree) to 7 (Strongly Agree)}} \\
BI1: Overall, would you use an SAV if given the chance? & \multirow{3}{*}{\cite{leeAutonomousVehiclesCan2019, daiImpactsIntroductionAutonomous2021}} \\
BI2: Would you likely use SAVs repeatedly in the future based on your experience today? & \\
BI3: Do you want to interact with this/your SAV again? & \\
\bottomrule
\multicolumn{2}{l}{\textit{*Note:} item PE1 is measured from 1 (Not fun at all) to 7 (Extremely fun)} \\
\end{tabular}
\end{adjustbox}
}
\vspace{-4mm}
\end{table*}

\subsection{Data availability}\label{sec: data availability}

The dataset collected and analyzed in this study has been made publicly available through an open-source repository \cite{guo2025sentiment}, as part of a comparative study on sentiment analysis. In contrast, the current manuscript presents an analysis of the survey responses accompanying this dataset, focusing on in-depth quantitative analyses and qualitative evaluations of psychological ownership, anthropomorphism, and user experiences in human-SAV interactions.

\subsection{Hypotheses and statistical analysis}

The collected data were analyzed using a combination of quantitative and qualitative methodologies. The quantitative analysis was conducted using the R statistical software (version 4.4.1), while the qualitative analysis was performed using the NVivo qualitative data analysis software (version 14.24.0).

\subsubsection{Descriptive statistics}

Means and standard deviations (Tables \ref{tab:mean_and_sd}) were calculated for all constructs within the survey to examine differences in participants' perceptions of the selected psychological factors and their UX. A bar chart (Fig. \ref{fig:pie}) was plotted to visualize the preferred SAV choices.

\subsubsection{Interaction type labeling and visualization}

To better characterize the diversity of user–SAV interactions, each conversational exchange (N = 2,136) was manually assigned a single task category (e.g., Navigation \& Motion Control, Entertainment Control, etc.; see Supplemental Materials Section~\ref{S-Appendix: interaction label} for the full list of labels and descriptions). Each user–agent exchange was converted to a vector representation using BERT word embeddings~\cite{devlinBERTPretrainingDeep2019}, as implemented via the Hugging Face transformers library~\cite{wolf-etal-2020-transformers} in Python. To visualize the distribution of interaction types, the high-dimensional embeddings were projected onto two dimensions using t-distributed stochastic neighbor embedding (t-SNE), implemented via the Rtsne package in R~\cite{Rtsne_package, Rtsne_paper} (Fig.~\ref{fig:t-SNE}).

\subsubsection{Favorite SAVs}

To investigate the effects of prompts on participants' UX with the four SAV types, we first tested whether participants' favorite SAVs varied significantly. If all SAVs were equally preferred, it could indicate that the tailored prompts utilized to develop the four SAV groups did not achieve the anticipated effect on users' experiences or choices. This resulted in the formulation of our initial hypothesis \textbf{H1}: The favorite SAVs are different across all participants.

The user study employed three distinct anthropomorphic characteristics for the SAVs, which were as follows: ``cool and sophisticated'', ``friendly and engaging'', and ``sassy and tired''. Participants selected their preferred anthropomorphic style before interacting with the SAVs and indicated their favorite SAV at the end of all interactions. It is important to ascertain whether participants' preferences for particular SAVs were shaped by the tailored prompts or by the SAVs' anthropomorphic attributes. In other words, participants may exhibit a systematic preference for one SAV over others based on their perception of its interaction style. This leads to our third hypothesis, which is as follows \textbf{H2}: There is an association between the selected anthropomorphism (e.g., friendly or cool) and the favorite SAV.

Non-parametric tests were used to test these hypotheses. Specifically, the Chi-Square Test and Fisher’s Exact Test were applied depending on whether the expected frequency for each category was greater or less than 5, respectively.

\subsubsection{Reliability check and ordering effect}

Before analyzing how different prompting strategies influenced UX with the designed SAV conversational agents, the internal consistency of questionnaire items was verified using Cronbach’s alpha. Since the same 50 participants interacted sequentially with all four SAV agents, there was a risk of carry-over effects, where participants’ evaluations might have been influenced by the order in which they experienced each SAV. Although the presentation order of SAVs was randomized to mitigate potential ordering biases, it remained important to explicitly test for such effects. For example, participants interacting with the later-presented SAV agents might report higher psychological ownership due to increasing familiarity, potentially altering the effectiveness of the prompts. To assess this potential ordering effect, Fisher’s Exact Test was employed to evaluate the relationship between the SAV presentation order and participants’ favorite SAV. To assess potential carry‐over effects on the remaining user‐experience constructs, we fitted a Linear Mixed‐Effects Model (LMM) that includes both which SAV participant interacted and presentation order, along with their interaction, while controlling for participant‐level variability via random intercepts \cite{meteyardBestPracticeGuidance2020}.  The model is specified in Equation \ref{eq:LMM_full}, where \(y_{ij}\) is the composite score (item average) for participant \(i\) under SAV condition \(j\) at presentation order \(j\), \(\mathrm{SAV}_{ij}\in\{\mathrm{SAV1},\dots,\mathrm{SAV4}\}\) is the SAV condition, \(\mathrm{Position}_{ij}\in\{1,2,3,4\}\) is the presentation position, \(\alpha_i\sim\mathcal{N}(0,\sigma_{\alpha}^2)\) is the participant‐specific random intercept, and \(\varepsilon_{ij}\sim\mathcal{N}(0,\sigma_{\varepsilon}^2)\) is the residual error \cite{schielzethRobustnessLinearMixedeffects2020}. The analysis was conducted using the R package `lme4' \cite{batesLme4LinearMixedeffects2025}.

\vspace{-4mm}
\begin{align}\label{eq:LMM_full}
y_{ij} &= \beta_0 
       + \beta_1\,\mathrm{SAV}_{ij}
       + \beta_2\,\mathrm{Position}_{ij}
       + \beta_3\,(\mathrm{SAV} \times \mathrm{Position})_{ij} \notag \\
&\quad + \alpha_i + \varepsilon_{ij}, \notag \\
\alpha_i &\sim \mathcal{N}(0,\sigma_{\alpha}^2), \quad
\varepsilon_{ij} \sim \mathcal{N}(0,\sigma_{\varepsilon}^2).
\end{align}

\subsubsection{Prompt effect}

The effects of the designed prompts were examined by comparing the survey responses collected after each interaction with the SAVs. The Friedman test, a non-parametric alternative to repeated measures ANOVA that is robust to departures from normality \cite{Nonparametric2015}, was conducted to determine whether there were significant differences between the four SAVs. Specifically, the test was applied to psychological ownership (PO), anthropomorphism (A), quality of service (QoS), disclosure tendency (DT), polarity (Pol), subjectivity (Sub), enjoyment of interaction (PE), and behavioral intention to use (BI), following verification of internal consistency using Cronbach’s alpha. Any significant variation in these constructs would indicate that the prompts influenced participants' perceptions and experiences, consistent with the following overarching hypothesis:

\begin{itemize}
    \item \textbf{H3}: A significant difference exists among the four SAV groups across the measured variables.
\end{itemize}

\begin{table}
\centering
\begin{threeparttable}
\caption{Summary of hypotheses tested in this study.}
\label{tab:hypotheses_methods}
\begin{tabularx}{\columnwidth}{@{}l >{\raggedright\arraybackslash}X@{}}
\toprule
ID & Hypothesis Statement \\
\midrule
H4.1 & SAV2 $>$ SAV1 on PO items (PO1–PO4). \\
H4.2 & SAV4 $>$ SAV3 on PO items (PO1–PO4). \\
H5.1 & SAV2 $>$ SAV1 on all UX metrics. \\
H5.2 & SAV4 $>$ SAV3 on all UX metrics. \\
H6.1 & SAV3 $>$ SAV1 on A items (A1–A5). \\
H6.2 & SAV4 $>$ SAV2 on A items (A1–A5). \\
H7.1 & SAV3 $>$ SAV1 on all UX metrics. \\
H7.2 & SAV4 $>$ SAV2 on all UX metrics. \\
H8.1 & SAV4 $>$ SAV1 on PO items (PO1–PO4). \\
H8.2 & SAV4 $>$ SAV1 on A items (A1–A5). \\
H8.3 & SAV4 $>$ SAV1 on all UX metrics. \\
\bottomrule
\end{tabularx}
\begin{tablenotes}[flushleft]\footnotesize
\item \textit{Notes:} SAV1 = Baseline; SAV2 = SAV1 + PO prompts; SAV3 = SAV1 + A prompts; SAV4 = SAV1 + PO and A prompts. 
UX = user-experience metrics, which include QoS, DT, sentiment of SAV responses (Polarity and Subjectivity), PE, BI.
\end{tablenotes}
\end{threeparttable}
\vspace{-4mm}
\end{table}

Based on prior literature (Sections~\ref{Lit:PO} and~\ref{Lit: A}), a set of hypotheses was developed to examine the effects of PO prompts, anthropomorphism prompts (A prompts), and their combination. These hypotheses are summarized in Table~\ref{tab:hypotheses_methods}. 
Hypotheses H4--H8 were tested at the item level using one-tailed Wilcoxon Signed-Rank tests, with Holm–Bonferroni corrections applied to control the family-wise error rate \cite{armstrongWhenUseBonferroni2014}.

\subsubsection{Qualitative analysis}
From the open-ended responses and interview data, seven themes emerged as drivers or barriers to the development of psychological ownership during interaction with the SAVs. Each theme was further broken down into specific codes reflecting participants' identified factors. The themes and codes are summarized in Table~\ref{tab:po_themes}.

\begin{table}[h]
\centering
\caption{Themes and codes influencing psychological ownership during SAV interactions.}
\label{tab:po_themes}
\begin{tabularx}{\columnwidth}{@{}p{3cm}X@{}}
\toprule
\textbf{Theme} & \textbf{Codes} \\
\midrule
Communication Style & Short and concise; Long and detailed; Interactive and engaging; Verbose and repetitive; Guidance and recommendations; Use of ``I''; Use of ``we'' \\
Communication Tone  & Professional, polite, and formal; Kind and warm; Neutral \\
Anthropomorphism    & Human-like; AI-/GPT-like; Robot-like; Machine-like \\
Performance         & Meeting user needs; Efficiency; Care and support; Accuracy; Intelligence and conversation memory; Perceived ease of use; Taking initial actions \\
Psychological Ownership Routes & Personalization; Familiarity, intimacy, and privacy; Feeling of control; Time and number of interactions \\
Legal Ownership     & Fact of shared car; Like a taxi driver/service \\
Sentiment           & Objective \\
\bottomrule
\end{tabularx}
\end{table}

The effects of each code on participants’ perceptions of psychological ownership were analyzed, noting whether each influence was positive or negative, and the frequency of occurrences was recorded. The descriptions and sample responses illustrating each code are provided in Table \ref{S-tab:prompt_code_summary} in the Supplemental Materials.

\section{Results}\label{sec: result}

\subsection{Descriptive Summary}

\begin{figure*}
    \centering
    \includegraphics[width=1.7\columnwidth]{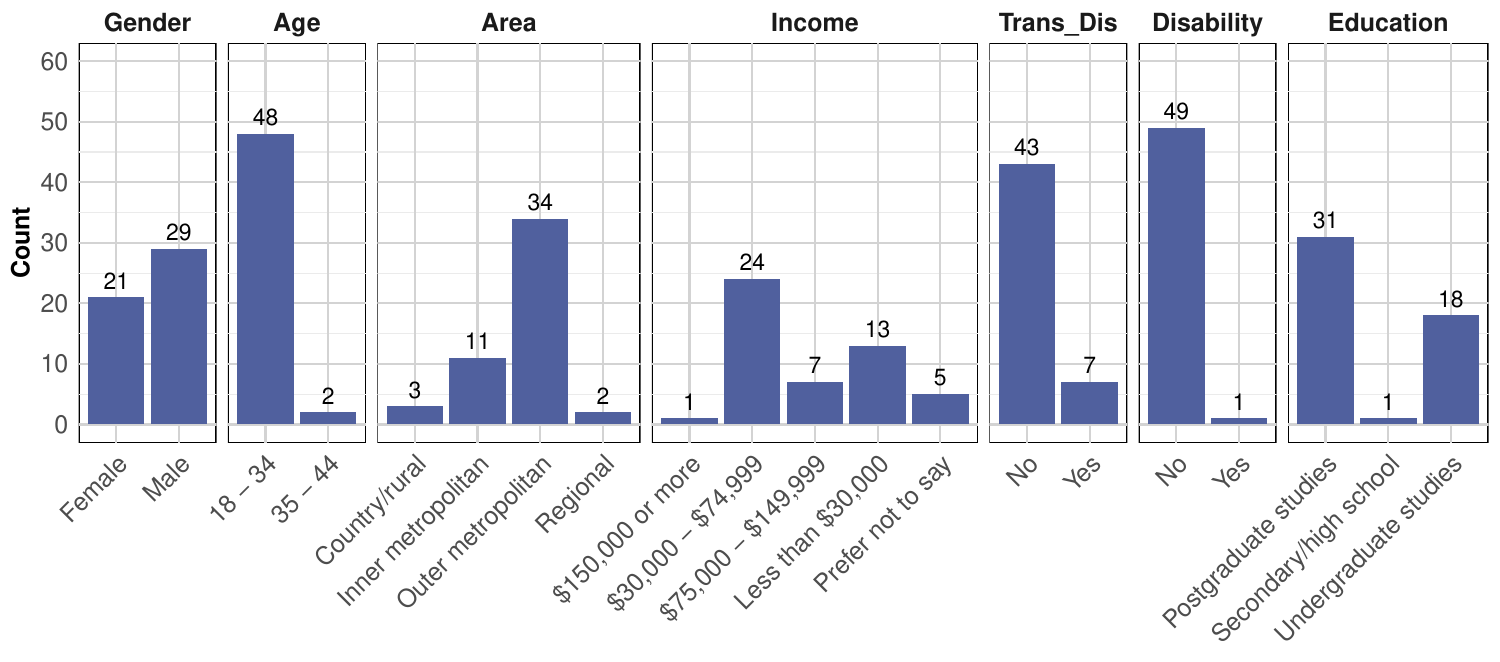}
    \caption{Distribution of demography (N = 50).}
    \label{fig:demographic}
    \vspace{-4mm}
\end{figure*}

The demographic profile of the respondents (Figure \ref{fig:demographic}) shows a relatively young, highly educated, and urban sample. 58\% of participants identified as male (n = 29), while 42\% identified as female (n = 21).  Most participants (86\%) reported not being responsible for transporting anyone with special care needs (e.g., elderly or disabled), while one participant identified as having a disability. Although this cohort is our main focus group, it should be noted that this demographic skew may limit the generalizability of our findings to the broader population, but highlights the segment likely to be the initial SAV market adopters.

\begin{figure}
    \centering
    \includegraphics[width=1\linewidth]{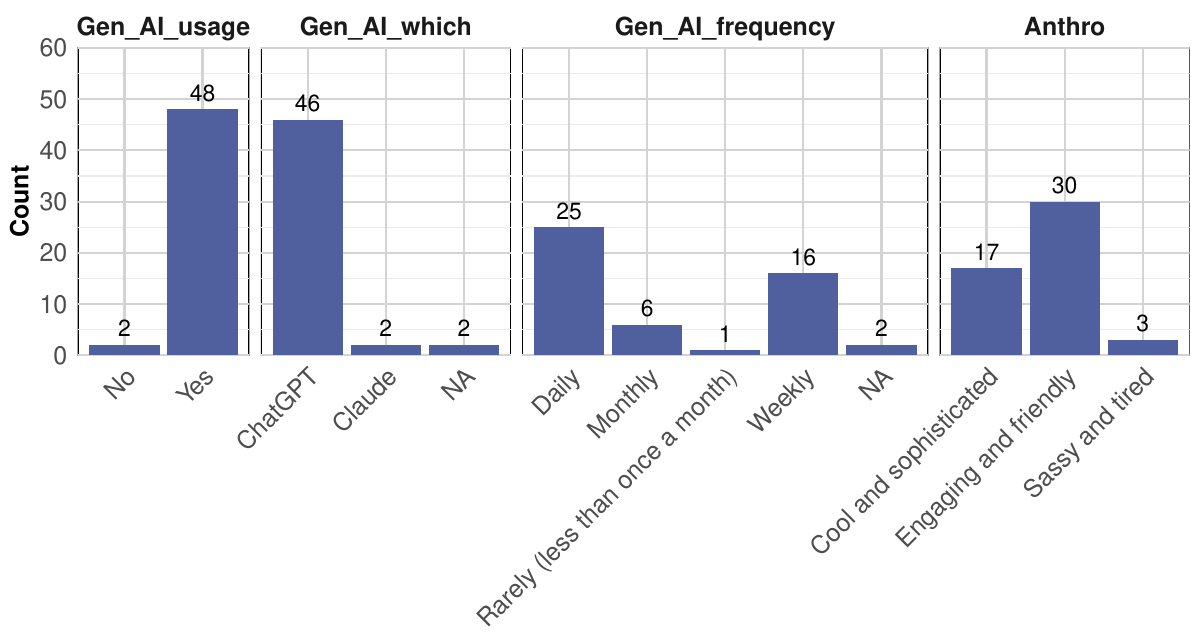}
    \caption{Distribution of AI usage (N = 50).}
    \label{fig:ai_use}
    \vspace{-4mm}
\end{figure}

Most of the participants (96\%) had used generative AI before, with ChatGPT being the most popular choice (92\%). In terms of usage frequency, half of the participants (50\%) reported using generative AI daily, followed by 32\% who used it weekly, 12\% monthly, and 2\% rarely (less than once a month). When selecting the anthropomorphic personas without any additional information, the ``engaging and friendly'' persona was the most preferred (60\%), followed by ``cool and sophisticated'' (34\%), while the ``sassy and tired'' persona was least favored (6\%). These results (Figure \ref{fig:ai_use}) reflect a high level of familiarity and frequent interaction with generative AI, alongside a preference for approachable and engaging anthropomorphic designs when participants are invited to interact with new SAVs.

Table \ref{table:mean} summarizes the mean values for each survey item across the four SAV groups, with the highest mean value for each item highlighted in bold. The results indicate that SAV4, which integrates both psychological ownership and anthropomorphism prompts, received the highest average ratings from participants. Specifically, participants rated SAV4 the highest in terms of their feelings of psychological ownership, perceived anthropomorphic qualities (cognitive, personality, emotional, and empathetic attributes), quality of service, willingness to share personal information, perceived positivity and subjectivity of the SAV’s responses, enjoyment of interaction, and behavioral intention to use the SAV.

Interestingly, a comparison between SAV1 (the control group) and SAV2 (the group with psychological ownership prompts) reveals that SAV2 shows higher mean values for PO1 (``I feel like the vehicle is mine'') and PO4 (``I feel like I own the vehicle''). However, it has lower mean values for PO2 (``I feel a very high degree of personal ownership of the vehicle'') and PO3 (``Interacting with the SAV interface enhances my feeling of ownership over the vehicle'') within the psychological ownership perspective. In contrast, the anthropomorphism prompts appear to be effective. SAV3 (the group with anthropomorphism personalities) demonstrates higher mean values for anthropomorphism items (A1--A5) compared to SAV1 and SAV2.

\begin{table*}[!t]
\centering
\caption{Mean and standard deviation survey results for each SAV group. The highest mean or standard deviation value for each survey item across the four groups is highlighted in bold.}
\label{tab:mean_and_sd}

\subfloat[Mean survey results for each group.\label{table:mean}]{
\resizebox{\textwidth}{!}{%
\begin{tabular}{lrrrrrrrrrrrrrrrrrrrrrr}
\toprule
SAV & PO1 & PO2 & PO3 & PO4 & A1 & A2 & A3 & A4 & A5 & QoS1 & QoS2 & QoS3 & DT1 & DT2 & DT3 & Pol1 & Sub & PE1 & PE2 & BI1 & BI2 & BI3\\
\midrule
1 & 3.56 & 3.64 & 4.10 & 3.38 & 3.78 & 3.32 & 3.22 & 2.80 & 3.44 & 5.22 & 5.12 & 5.12 & 4.88 & 5.46 & 4.24 & 0.41 & 0.35 & 4.18 & 5.00 & 5.58 & 5.14 & 5.02\\
2 & 3.70 & 3.56 & 3.90 & 3.46 & 3.86 & 3.62 & 3.36 & 3.10 & 3.48 & 5.44 & 5.28 & 5.48 & 4.96 & 5.78 & 4.20 & 0.43 & 0.36 & 4.42 & 5.22 & 5.68 & 5.32 & 5.16\\
3 & 3.66 & 3.48 & 3.86 & 3.36 & 4.24 & 3.98 & 3.60 & 3.24 & 3.58 & 5.28 & 5.12 & 5.28 & 4.88 & 5.54 & 4.44 & 0.45 & 0.38 & 4.54 & 5.24 & 5.60 & 5.24 & 5.08\\
4 & \textbf{4.14} & \textbf{4.04} & \textbf{4.48} & \textbf{3.96} & \textbf{4.48} & \textbf{4.30} & \textbf{3.92} & \textbf{3.64} & \textbf{3.98} & \textbf{5.70} & \textbf{5.50} & \textbf{5.54} & \textbf{5.06} & \textbf{5.86} & \textbf{4.62} & \textbf{0.62} & \textbf{0.49} & \textbf{4.92} & \textbf{5.34} & \textbf{5.84} & \textbf{5.60} & \textbf{5.32}\\
\bottomrule
\end{tabular}}
}

\vspace{-0.5em} 

\subfloat[Standard deviation survey results for each group.\label{table:sd}]{
\resizebox{\textwidth}{!}{%
\begin{tabular}{lrrrrrrrrrrrrrrrrrrrrrr}
\toprule
SAV & PO1 & PO2 & PO3 & PO4 & A1 & A2 & A3 & A4 & A5 & QoS1 & QoS2 & QoS3 & DT1 & DT2 & DT3 & Pol1 & Sub & PE1 & PE2 & BI1 & BI2 & BI3\\
\midrule
1 & 1.69 & 1.54 & 1.58 & 1.69 & 1.68 & 1.73 & 1.79 & 1.58 & 1.75 & 1.37 & 1.41 & 1.45 & 1.73 & 1.39 & 1.86 & \textbf{0.46} & 0.28 & \textbf{1.73} & \textbf{1.75} & 1.34 & 1.37 & 1.45\\
2 & 1.74 & 1.66 & 1.74 & 1.72 & 1.86 & 1.79 & \textbf{1.87} & 1.83 & 1.81 & 1.25 & 1.28 & 1.16 & 1.76 & 1.17 & 1.90 & 0.46 & 0.26 & 1.50 & 1.61 & \textbf{1.38} & 1.41 & \textbf{1.62}\\
3 & 1.71 & 1.72 & 1.73 & 1.65 & 1.96 & 1.73 & 1.69 & 1.72 & 1.82 & 1.28 & 1.41 & 1.46 & 1.80 & \textbf{1.47} & 1.89 & 0.44 & 0.27 & 1.53 & 1.66 & 1.23 & \textbf{1.42} & 1.48\\
4 & \textbf{1.95} & \textbf{1.97} & \textbf{1.84} & \textbf{2.00} & \textbf{1.99} & \textbf{1.87} & 1.86 & \textbf{2.01} & \textbf{1.97} & \textbf{1.46} & \textbf{1.56} & \textbf{1.55} & \textbf{1.81} & 1.25 & \textbf{1.94} & 0.37 & \textbf{0.29} & 1.50 & 1.44 & 1.30 & 1.34 & 1.54\\
\bottomrule
\end{tabular}}
}
\end{table*}

The mean values for each measured factor, categorized by participants’ preferred SAV, are presented in Table~\ref{S-tab:mean_favor} in the Supplemental Materials. These results demonstrate that participants consistently rated their favored SAV significantly higher across nearly all metrics. In other words, when a specific SAV was chosen as a favorite, it received relatively higher ratings in psychological ownership, perceived anthropomorphism, service quality, willingness to disclose personal information, sentiment of SAV responses, interaction enjoyment, and behavioral intention to use the SAV.

\subsection{Interaction type labeling and visualization}

To better understand the diversity of user–SAV interactions, each conversational exchange was manually labeled with one of eleven task categories (see Table~\ref{S-tab:interaction_label} in the Supplemental Materials). Figure~\ref{fig:t-SNE} visualizes the distribution of these interaction types using t-SNE, based on BERT embeddings of each exchange. The resulting plot reveals clear clustering by task category. The most frequent interaction types were \textit{Navigation \& Motion Control} (468 instances), \textit{Cabin Environment} (426), and \textit{Entertainment Control} (345), followed by \textit{System Capabilities \& Policies} (170), \textit{External Communication} (143), and \textit{Well-being Support} (134). Notably, 69 interactions focused on personalization requests such as creating or loading personal profiles, while 55 exchanges pertained to local context queries, such as seeking recommendations for nearby restaurants.

\begin{figure}
    \centering
    \includegraphics[width=1\linewidth]{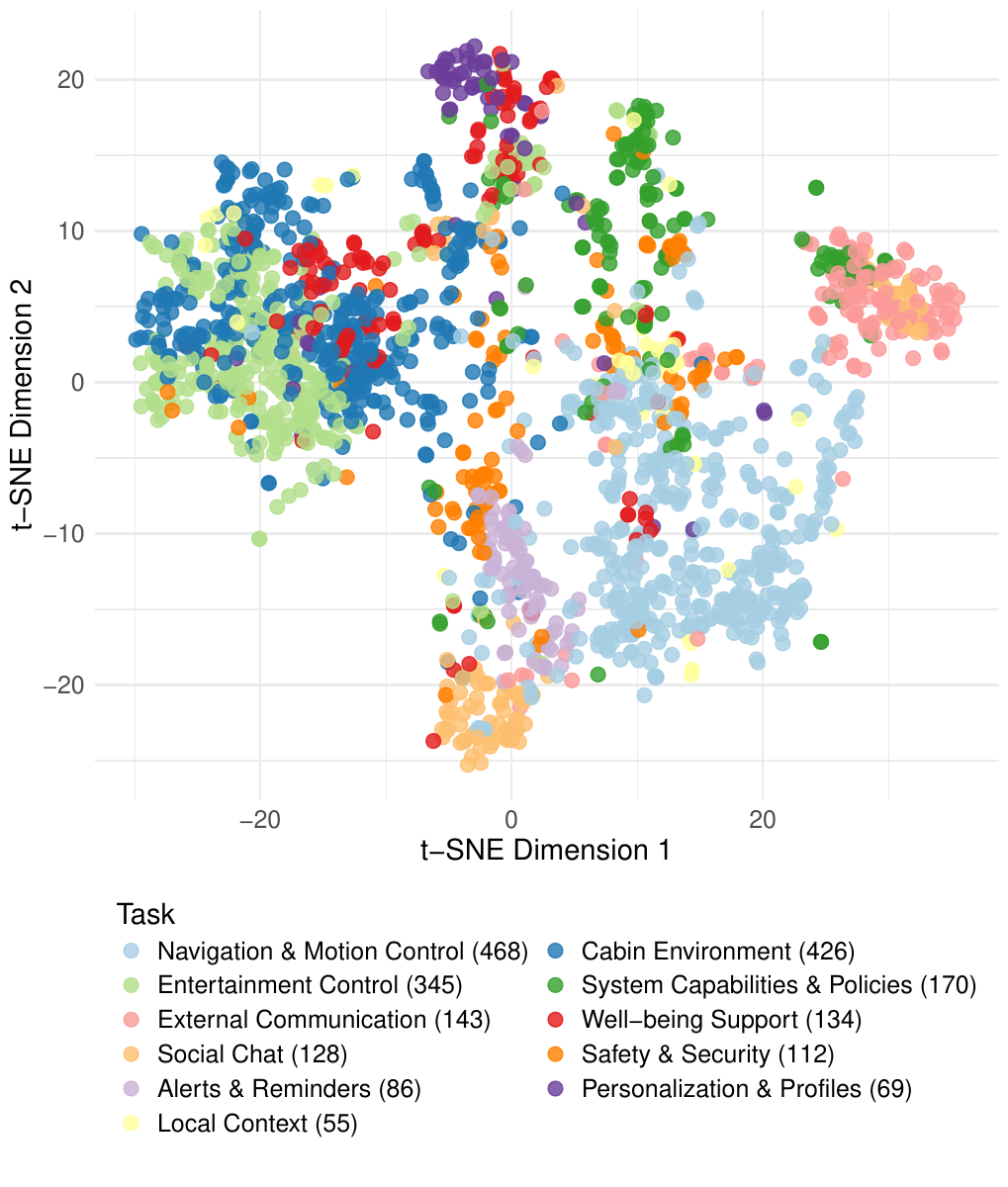}
    \caption{2D t-SNE Visualization of SAV Conversations Text Embeddings Colored by Task Type, ranked by repeated counts.}
    \label{fig:t-SNE}
    \vspace{-4mm}
\end{figure}

\begin{figure}
    \centering
    \includegraphics[width=0.8\linewidth]{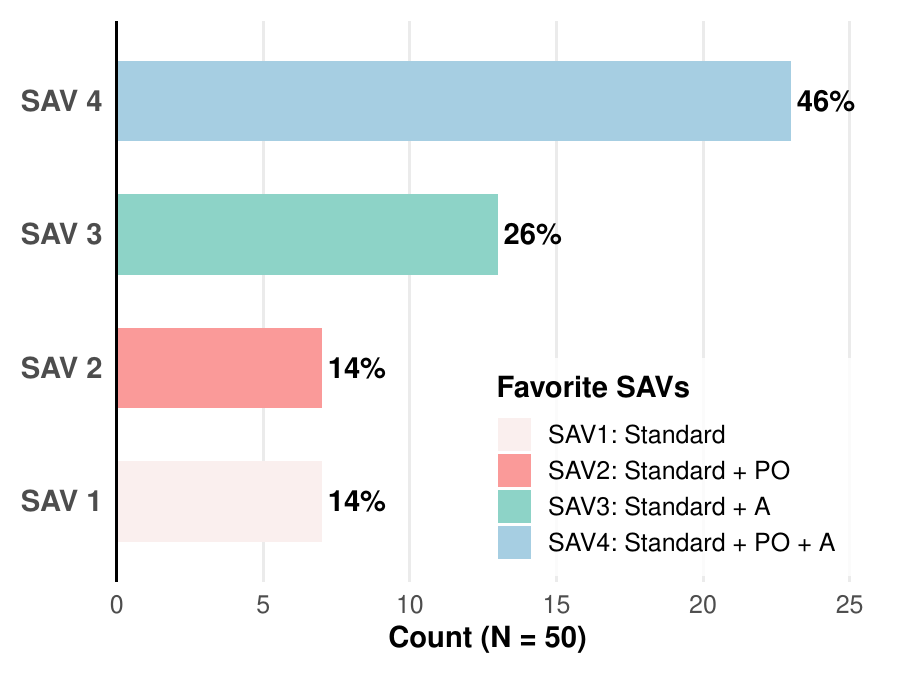}
    \caption{Distribution of participants' favorite SAV selections across the four SAV designs.}
    \label{fig:pie}
    \vspace{-4mm}
\end{figure}

\subsection{Favorite SAVs}

Among the four designed SAVs, SAV4 emerged as the most favored SAV (Figure \ref{fig:pie}, selected by 46\% of participants after interacting with all SAVs in random order. This result aligns with the mean survey findings in Table \ref{table:mean}, where SAV4 achieved the highest average scores across all survey items. 

\begin{table}[!t]
\centering
\caption{Hypothesis Testing Results. Significance levels: *** $p<0.001$, ** $p<0.01$, * $p<0.05$.}
\label{tab:hypothesis_results}
\renewcommand\tabularxcolumn[1]{>{\raggedright\arraybackslash}m{#1}}
\begin{tabularx}{\linewidth}{X l l}
\toprule
\textbf{Hypothesis} & \textbf{Test} & \textbf{p-value}\\ \midrule
\textbf{H1}: The favorite SAVs are different across all participants & Chi-Square Test & 0.0034**\\ 
\textbf{H2}: There is an association between the selected anthropomorphism (e.g., friendly or cool) and the favorite SAV. & Fisher’s Exact Test & 0.0524\\ 
\bottomrule
\end{tabularx}
\vspace{-4mm}
\end{table}

The Chi-Square Test and Fisher's Exact Test were used to evaluate \textbf{H1} and \textbf{H2}, with the results summarized in Table \ref{tab:hypothesis_results}. Among the tested hypotheses, only \textbf{H1} passed the significance threshold (p-value $<$ 0.05). This finding suggests that participants' favorite SAVs differ among the four developed SAV styles, indicating that the designed prompts effectively produce distinct SAV characteristics. No significant evidence was found to support a relationship between the favorite SAV order and the favorite SAV, nor between the selected anthropomorphism style and the favorite SAV.

\subsection{Reliability check and ordering effect}

Table \ref{tab:cronbach_alpha} presents the Cronbach’s alpha values for the measured factors, which indicate the internal consistency and reliability of the scales adopted. All factors except perceived enjoyment achieved a Cronbach’s alpha value above 0.7, which indicates that the measures and constructs were acceptable in terms of the internal reliability \cite{UCLA_cronbachs}. Since perceived enjoyment (\(alpha\) = 0.5795) falls below the commonly accepted threshold, suggesting that the scale may lack sufficient internal consistency, it is excluded from further analysis.

\begin{table}[ht]
\centering
\caption{Cronbach's Alpha Values for Measured Factors.}
\label{tab:cronbach_alpha}
\begin{tabular}{@{}lc@{}}
\toprule
\textbf{Factor}               & \textbf{Raw Alpha} \\ \midrule
Psychological Ownership       & 0.9588             \\
Anthropomorphism              & 0.9320             \\
Quality of Service            & 0.9303             \\
Disclosure Tendency           & 0.7601             \\
Perceived Enjoyment           & 0.5795             \\
Behavioral Intention          & 0.9059             \\ \bottomrule
\end{tabular}
\vspace{-4mm}
\end{table}

Fisher’s Exact Test indicated no statistically significant association between SAV presentation order and participants’ favorite choice (\(p = 0.145\)), suggesting no ordering bias.

Prior to conducting the Linear Mixed-Effects Model analyses, key model assumptions were carefully examined. The Shapiro–Wilk tests and visual inspection of QQ plots indicated that the residuals for Psychological Ownership and Anthropomorphism were approximately normally distributed (\(p > 0.05\)). However, residuals for Quality of Service, Disclosure Tendency, Polarity, Subjectivity, and Behavioral Intention significantly deviated from normality. Nonetheless, considering the sufficiently large sample size (\(N = 200\)) and the robustness of the Linear Mixed-Effects Models against moderate departures from normality assumptions \cite{schielzethRobustnessLinearMixedeffects2020}, the use of the Linear Mixed-Effects Models remains appropriate. 

Results from Type-III ANOVA conducted using LMM (Table~\ref{tab:lmm_anova}) indicate significant main effects of SAV conditions on Anthropomorphism, Polarity, and Subjectivity, while no significant SAV effects were observed for other psychological factors. Position effects, representing potential ordering or carry-over effects, were not statistically significant (\(p>0.05\)) for any of the investigated constructs. Additionally, interactions between the SAV\(\times\)Position were non-significant across all measured factors. To further assess potential ordering effects, the Friedman test was conducted for each psychological factor, followed by pairwise Wilcoxon signed-rank tests comparing each SAV presentation position. Holm–Bonferroni correction was applied to control for multiple comparisons. None of the pairwise comparisons yielded adjusted p-values below 0.05. These findings collectively suggest minimal carry-over effects due to the sequential SAV presentations, indicating that the observed significant differences are primarily attributable to the distinct characteristics of the SAVs themselves, rather than their presentation order.

\begin{table*}[!h]
\centering
\begin{threeparttable}
\caption{Type-III ANOVA for SAV condition, Position, and their interaction from LMM analysis.}
\label{tab:lmm_anova}
\begin{tabular}{p{3cm} p{2cm} p{1.5cm} c c c c}
\toprule
Outcome                &
\multicolumn{1}{c}{SAV F (df)} & \multicolumn{1}{c}{p-value} &
\multicolumn{1}{c}{Position F (df)} & \multicolumn{1}{c}{p-value} &
\multicolumn{1}{c}{SAV×Position F (df)} & \multicolumn{1}{c}{p-value} \\
\midrule
Psychological Ownership & 1.59 (3,149.3) & 0.194 & 0.94 (3,149.3) & 0.422 & 0.36 (9,178.2) & 0.954\\
Anthropomorphism        & 5.37 (3,149.9) & 0.002** & 1.56 (3,149.9) & 0.201 & 0.58 (9,163.6) & 0.809\\
Quality of Service      & 1.75 (3,152.3) & 0.159 & 1.23 (3,152.3) & 0.299 & 1.19 (9,198.7) & 0.300\\
Disclosure Tendency     & 1.13 (3,150.5) & 0.339 & 1.81 (3,150.5) & 0.148 & 1.48 (9,163.0) & 0.161\\
Polarity                & 3.84 (3,149.5) & 0.011* & 2.25 (3,149.5) & 0.085 & 0.96 (9,196.4) & 0.474\\
Subjectivity            & 4.30 (3,150.5) & 0.006** & 0.61 (3,150.5) & 0.611 & 0.44 (9,179.3) & 0.911\\
Behavioral Intention    & 0.87 (3,150.8) & 0.461 & 0.26 (3,150.7) & 0.857 & 1.18 (9,195.9) & 0.308\\
\bottomrule
  \end{tabular}
  \begin{tablenotes}
    \footnotesize\raggedright
    \item Note. df pairs are (df$_1$, df$_2$), where df$_1$ is the numerator degrees of freedom (levels–1 = 3) and df$_2$ is the denominator df estimated via the Satterthwaite approximation. * $p<.05$; ** $p<.01$.
  \end{tablenotes}
\end{threeparttable}
\end{table*}

\subsection{Prompt effect}

Given the ordinal nature and non-normal distribution of the item-level data, the Friedman test was employed to assess significant differences among the four dependent groups. Table~\ref{tab:Friedman_results} presents items with significant differences (\(p<0.05\)). While statistically significant, the effect sizes are relatively small (values below 0.3) \cite{kassambaraPipeFriendlyFrameworkBasic2023}. 

Notably, all five Anthropomorphism items showed significant differences according to the Friedman test. Additionally, participants’ willingness to disclose personal feelings (DT3) while interacting with the SAVs, their perceived polarity (Pol) and subjectivity (Sub) of the SAV responses, and their intention to use SAVs repeatedly in the future (BI2) also demonstrated statistical significance. Thus, \textbf{H3} is only partially supported. The Friedman test results align closely with the LMM findings (Table~\ref{tab:lmm_anova}), which also showed significant differences in Anthropomorphism, Polarity, and Subjectivity (item average) across the SAVs. Due to widespread violation of normality at the item level, Friedman test results were prioritized in the subsequent discussion.


\begin{table*}[h]
\centering
\caption{Summary of Friedman test results for items with statistically significant p-values (\(p<0.05\)) using Kendall’s W as the effect size method. All tests have 3 degrees of freedom.}
\label{tab:Friedman_results}
\begin{tabular}{lp{10cm}rrrrll}
\toprule
\textbf{Item} & \textbf{Statement} & \textbf{Chi-Squared} & \textbf{p-value} & \textbf{Effect Size} & \textbf{Magnitude} \\
\midrule
A1 & \raggedright I believe that the SAV has its own decision-making power. & 13.767 & 0.0032 & 0.092 & Small \\
A2 & \raggedright I believe that the SAV has its own personality. & 17.435 & 0.0006 & 0.116 & Small \\
A3 & \raggedright I believe that the SAV has its own preferences and mood. & 11.295 & 0.0102 & 0.075 & Small \\
A4 & \raggedright I believe that the SAV has its own emotions. & 14.091 & 0.0028 & 0.094 & Small \\
A5 & \raggedright I believe that the SAV device can feel compassion. & 7.875 & 0.0487 & 0.053 & Small \\
DT3 & \raggedright I am willing to disclose my feelings, including negative emotions like anxiety, with the SAV. & 11.459 & 0.0095 & 0.076 & Small \\
Pol & \raggedright Overall, how do you rate the SAV responses, negative or positive? & 10.649 & 0.0138 & 0.071 & Small \\
Sub & \raggedright Overall, do you think the SAV responses are objective or subjective? & 12.600 & 0.0056 & 0.084 & Small \\
BI2 & \raggedright Would you likely use SAVs repeatedly in the future based on your experience today? & 9.157 & 0.0273 & 0.061 & Small \\
\bottomrule
\end{tabular}
\end{table*}

The one-tailed Wilcoxon Signed-Rank test results are summarized in Table \ref{tab:Pairwise_comparisons}, with significance determined based on p-adjusted values after applying Holm–Bonferroni corrections. Detailed results for each hypothesis are provided in Tables \ref{S-tab:H6_results}-\ref{S-tab:H9_results} in the Supplemental Materials Section~\ref{S-Appendix: Wilcoxon results}.

\begin{table*}[h]
\centering
\caption{Summary of pairwise comparisons for user perception and experience across SAV groups. A checkmark (\checkmark) denotes a statistically significant difference between the corresponding groups (e.g., SAV2 $>$ SAV1), while a cross (\textcolor{gray}{\(\times\)}) indicates no significant difference. Significance was determined based on p-adjusted values after applying Holm–Bonferroni corrections.}
\label{tab:Pairwise_comparisons}
\begin{tabular}{lccccc}
\toprule
\textbf{Item} & \textbf{SAV2 $>$ SAV1} & \textbf{SAV3 $>$ SAV1} & \textbf{SAV4 $>$ SAV1} & \textbf{SAV4 $>$ SAV2} & \textbf{SAV4 $>$ SAV3} \\
\midrule
A1     & --  & \textcolor{gray}{\(\times\)} & \checkmark & \checkmark & --  \\
A2     & --  & \checkmark & \checkmark & \checkmark & --  \\
A3     & --  & \textcolor{gray}{\(\times\)} & \checkmark & \checkmark & --  \\
A4     & --  & \checkmark & \checkmark & \checkmark & --  \\
A5     & --  & \textcolor{gray}{\(\times\)} & \textcolor{gray}{\(\times\)} & \checkmark & --  \\
DT3    & \textcolor{gray}{\(\times\)} & \textcolor{gray}{\(\times\)} & \textcolor{gray}{\(\times\)} & \textcolor{gray}{\(\times\)} & \textcolor{gray}{\(\times\)} \\
Pol    & \textcolor{gray}{\(\times\)} & \textcolor{gray}{\(\times\)} & \checkmark & \checkmark & \textcolor{gray}{\(\times\)} \\
Sub    & \textcolor{gray}{\(\times\)} & \textcolor{gray}{\(\times\)} & \checkmark & \checkmark & \checkmark \\
BI2    & \textcolor{gray}{\(\times\)} & \textcolor{gray}{\(\times\)} & \textcolor{gray}{\(\times\)} & \textcolor{gray}{\(\times\)} & \textcolor{gray}{\(\times\)} \\
\bottomrule
\end{tabular}
\vspace{-4mm}
\end{table*}

\subsubsection{Psychological ownership prompt effect}

Since none of the PO items were statistically significant under the Friedman Test, this indicates that there is no significant difference among the four SAV groups in participants' perceptions of ownership after interacting with the SAVs. Consequently, \textbf{H4.1} and \textbf{H4.2} are not supported. To better understand the mechanisms behind these results and how the prompts influenced participants’ sense of ownership, participants were interviewed immediately after each interaction to describe their feelings of ownership of that SAV. The qualitative findings from these interviews are discussed in Section \ref{Sec: PO_interview}.

When comparing SAV2 $>$ SAV1 and SAV4 $>$ SAV3 to examine the effects of PO prompts on UX items, only the perceived subjectivity (Sub) of the SAV responses demonstrated a significant directional difference. Specifically, participants indicated that SAV4’s responses were perceived as more subjective than SAV3’s responses, suggesting that PO prompts, in association with anthropomorphism prompts, may increase perceptions of subjectivity. In contrast, no significant difference was found when comparing SAV2 and SAV1. One possible explanation is that the presence of a certain degree of anthropomorphism (e.g., A prompts) may increase the influence of PO prompts on perceived subjectivity, making the SAV responses appear more subjective. Consequently, \textbf{H5.1} is not supported and \textbf{H5.2} is only partially supported.

\subsubsection{Anthropomorphism prompt effect}

With the anthropomorphism prompt integrated, SAV3 exhibited a stronger perceived personality (A2) and a greater capacity for emotion (A4) than SAV1, thus \textbf{H6.1} is partially supported. This outcome aligns with expectations, as the prompt explicitly detailed how the SAV should interact in a more human-like manner (e.g., ``embodying a friendly and engaging persona''). Notably, even a single sentence instructing the SAV to act more anthropomorphic was sufficient to manipulate users’ perceptions of the system’s human-like attributes. 

A similar pattern emerged when comparing SAV4 (with both PO and A prompts) to SAV2 (with only PO prompts). In every anthropomorphism measure (i.e., perceived decision-making power, personality, preferences and mood, emotional capacity, and empathy), SAV4 significantly outperformed SAV2, which shows that \textbf{H6.2} is supported. This result implies that layering anthropomorphic interaction styles on top of PO prompts produces a deeper transformation in how users perceive the human-like qualities of the SAV. 

Several factors may explain why SAV4 elicits stronger anthropomorphic attributions than SAV2, while also yielding differences in more aspects compared to SAV3 and SAV1. First, anthropomorphism prompts tend to be more direct and include explicit instructions (e.g., behave in a ``warm and approachable'' manner), making it easier for the LLM to shift its role accordingly and convey human-like characteristics to users. In contrast, psychological ownership prompts are inherently less direct and lack specific instructions such as on how to foster perceptions of control or personal investment. This ambiguity likely makes it harder for the LLM to induce a sense of ownership, instead making participants more likely to perceive human-like qualities (e.g., emotions, decision-making agency) in the SAV. In addition, anthropomorphism prompts may synergistically reinforce any slight ownership cues, prompting users to see a personal connection to the vehicle. When users feel a sense of personal connection to the vehicle, they may become more sensitive to SAV's human-like features. Consequently, the interplay of perceived ownership and explicit anthropomorphic cues appears to amplify users’ recognition of anthropomorphic traits compared to interactions that integrate only psychological ownership cues.

When examining the effects of A prompts on UX by comparing SAV3 and SAV1, no significant differences emerged, indicating that \textbf{H7.1} is not supported. However, in the comparison of SAV4 against SAV2, participants perceived a higher level of sentiment in the SAV4 responses. Specifically, they found these responses to be more positive and subjective than those of SAV2, thus providing partial support for \textbf{H7.2}.

These findings suggest that simply setting an interaction style (i.e., A prompts) does not necessarily alter users’ perceptions of service quality, SAV response sentiment, disclosure tendencies, or behavioral intention. Nevertheless, when combined with PO prompts, the addition of anthropomorphism cues appears to enhance users’ impressions of the SAV’s responses, making them seem more positive and subjective.

\subsubsection{Combined prompt effect}

When SAV4 was compared against SAV1 on the items that showed significance in the Friedman test, the combination of both prompts led to higher ratings for decision power (A1), personality (A2), preferences and mood (A3), emotions (A4), and the sentiment of the SAV’s responses (Pol, Sub). Consequently, \textbf{H8.1} was not supported, whereas \textbf{H8.2} and \textbf{H8.3} received partial support.

These findings reveal a noteworthy pattern. While SAV2 (PO prompts only) exhibited perceptions akin to SAV1, and SAV3 (A prompts only) increased ratings for personality (A2) and emotions (A4) without improving broader UX factors, SAV4 (combined prompts) enhanced a wider range of anthropomorphic traits and response sentiment. Specifically, participants perceived SAV4 to possess greater decision power and more defined preferences/moods, and they considered its responses to be more positive. However, they also perceived SAV4 responses to be more subjective. This indicates that the combined effect of PO and A prompts exceeds the sum of their individual contributions, consistent with comparisons of SAV4 vs. SAV2 and SAV4 vs. SAV3. These findings underscore the value of multi-faceted design strategies in conversational systems. While prompts targeting psychological ownership alone or anthropomorphism alone may produce modest gains in user perceptions, especially user experience, combining these elements appears to result in a more pronounced response.

The results of prompt testing hypotheses are summarized in Table \ref{tab:hypothesis_results_5-9}.

\begin{table}
\centering
\caption{Results of one-tailed Wilcoxon Signed-Rank tests with Holm–Bonferroni adjustment.}
\label{tab:hypothesis_results_5-9}
\begin{tabular}{p{4.7cm}p{1cm}p{2cm}}
\toprule
Hypothesis Statement & Outcome & Significant Items \\
\midrule
\textbf{H4.1}: SAV2 $>$ SAV1 on PO1–PO4. & No & None \\
\textbf{H4.2}: SAV4 $>$ SAV3 on PO1–PO4. & No & None \\
\textbf{H5.1}: SAV2 $>$ SAV1 on all UX metrics. & No & None \\
\textbf{H5.2}: SAV4 $>$ SAV3 on all UX metrics. & Partially & Sub \\
\textbf{H6.1}: SAV3 $>$ SAV1 on A1–A5. & Partially & A2, A4 \\
\textbf{H6.2}: SAV4 $>$ SAV2 on A1–A5. & Yes & A1--A5 \\
\textbf{H7.1}: SAV3 $>$ SAV1 on all UX metrics. & No & None \\
\textbf{H7.2}: SAV4 $>$ SAV2 on all UX metrics. & Partially & Pol, Sub \\
\textbf{H8.1}: SAV4 $>$ SAV1 on PO1–PO4. & No & None \\
\textbf{H8.2}: SAV4 $>$ SAV1 on A1–A5. & Partially & A1--A4 \\
\textbf{H8.3}: SAV4 $>$ SAV1 on all UX metrics. & Partially & Pol, Sub \\
\bottomrule
\end{tabular}
\vspace{-4mm}
\end{table}

\subsection{Psychological ownership -- Insights from open-ended questions and interview responses}\label{Sec: PO_interview}

Although statistical tests revealed no significant differences in participants' perceived psychological ownership of the SAVs they interacted with across the four SAV groups, insights from interviews and responses to open-ended questions painted a more nuanced picture.  Despite knowing that the SAV was not theirs, a variety of perspectives emerged when participants were asked whether they perceived a sense of psychological ownership of the SAV they had just interacted with; any changes from the previous one(s); and responded to an open-ended question about additional features or experiences that might have enhanced their sense of ownership. The responses were grouped into several themes, as summarized in Table \ref{S-tab:prompt_code_summary} in the Supplemental Materials.

\begin{table*}[!ht]
\centering
\caption{Top 15 Codes Mentioned by the Participants that Affect Psychological Ownership in SAV Interactions. The order is ranked by No. of Unique People mentioned the codes. No. mentioned counts the total mentions of that code.}
\label{tab:15codes}
\begin{tabular}{@{}cllcccc@{}}
\toprule
\textbf{Order} & \textbf{Themes} & \textbf{Codes} & \textbf{Effects} & \textbf{No. of Unique People} & \textbf{No. mentioned} \\ 
\midrule
1  & Performance                     & Meeting user's needs                      & (+) & 25 & 49 \\
2  & Performance                     & Intelligence and conversation memory      & (+) & 22 & 36 \\
3  & Psychological Ownership Routes  & Familiarity, intimacy, and privacy        & (+) & 22 & 53 \\
4  & Psychological Ownership Routes  & Personalization                           & (+) & 21 & 39 \\
5  & Performance                     & Care and supportive                       & (+) & 16 & 27 \\
6  & Communication Style             & Interactive and engaging                  & (+) & 15 & 25 \\
7  & Legal Ownership                 & Like a taxi driver/service                & (-) & 13 & 20 \\
8  & Anthropomorphism                & Human-like                                & (+) & 12 & 18 \\
9  & Communication Style             & Verbose and repetitive                    & (-) & 10 & 12 \\
10 & Performance                     & Efficiency                                & (+) & 10 & 16 \\
11 & Psychological Ownership Routes  & Feeling of control                        & (+) & 10 & 13 \\
12 & Psychological Ownership Routes  & Time and number of interactions           & (+) & 9  & 13 \\
13 & Communication Tone              & Professional, polite, and formal          & (-) & 9  & 9  \\
14 & Anthropomorphism                & Machine-like                              & (-) & 8  & 9  \\
15 & Performance                     & Accuracy                                  & (+) & 8  & 12 \\
\bottomrule
\end{tabular}
\vspace{-4mm}
\end{table*}

\subsubsection{Top 15 codes}

After analyzing the responses, the top 15 most frequently mentioned themes are summarized in Table \ref{tab:15codes}. Notably, although participants were not explicitly informed about the psychological ownership theory or its underlying mechanisms, their textual responses validate key psychological ownership routes. All four codes of the \textbf{Psychological Ownership Routes} theme appeared in the responses, including \textit{Familiarity, intimacy, and privacy} (i.e., intimate knowledge, mentioned by 22 participants), \textit{Personalization} (i.e., control and investment of oneself, mentioned by 16 participants), \textit{Feeling of control} (i.e., control, mentioned by 10 participants), and \textit{Time and number of interactions} (related to intimate knowledge and investment of oneself, mentioned by 9 participants). These findings strongly support the psychological ownership framework articulated by \cite{pierceHistoryPsychologicalOwnership2018} in the context of SAVs, indicating that the three routes (i.e., intimate knowledge, control, and self-investment) can indeed be leveraged to enhance users’ sense of ownership during human-SAV interactions.

Beyond the three routes proposed by \cite{pierceHistoryPsychologicalOwnership2018}, our qualitative analysis identified a fourth emergent theme, perceived Performance, related to participants' perceptions of the conversational SAV agents' accuracy, efficiency, and intelligence. \textbf{Performance} was notably the theme mentioned the most frequently among the top 15, with associated codes such as \textit{Meeting user’s needs} (mentioned by 25 participants), \textit{Intelligence and conversation memory} (22 participants), \textit{Care and supportive} (16 participants), \textit{Efficiency} (10 participants), and \textit{Accuracy} (8 participants). Participants consistently indicated that a vehicle failing to meet their functional needs would not be perceived as truly ``theirs''. Conversely, participants reported a stronger sense of ownership when the SAV demonstrated consistent comprehension of user requests, provided intelligent and supportive responses, and performed tasks efficiently. These findings suggest that perceived performance of the conversational SAV agents may be an important factor affecting psychological ownership within highly automated mobility contexts. This interpretation aligns with \cite{murugananthamGratificationdrivenCustomerEngagement}, who identified performance expectancy (i.e., the degree to which users perceive technology as enhancing their efficiency) as a mediator between customer engagement and psychological ownership in the context of AR mobile applications.
However, this additional pathway requires further empirical validation. Although based on a single SAV study and requiring further empirical validation, these results point to a promising direction for expanding and refining psychological ownership frameworks in AI-driven transportation and agents.

Collectively, these findings suggest that although participants did not express strong psychological ownership in response to the four direct survey questions, their qualitative feedback indicates that psychological ownership can be fostered by improving SAV performance and more effectively fulfilling the key psychological ownership routes. Future SAV designs should prioritize these aspects to enhance users’ sense of psychological ownership in human-SAV interactions.

\subsubsection{Codes positively affect psychological ownership}

All codes within the \textbf{Performance} and \textbf{Psychological Ownership Routes} themes were perceived to have positive effects on fostering psychological ownership during interactions with SAVs. Additionally, other positively perceived factors included an \textit{interactive and engaging} communication style (mentioned by 22 participants), \textit{providing guidance and recommendations} during interactions (6 participants), and using a \textit{kind and warm tone} (6 participants).

An especially interesting insight, though mentioned by only one participant, was the impact of using inclusive language such as ``we''. This participant stated, ``It is caring and uses `we', which makes me feel like it is mine.'' and noted the opposite effect when the SAV used language framed in the first person, saying, ``The responses were all written in a style like `I will...,' which makes it feel like the AI owns the vehicle, but not me.'' This observation aligns with findings from \cite{peckCaringCommonsUsing2021}, who stated that strategically framed language (e.g., ``welcome to YOUR park'') can strengthen psychological ownership.

\subsubsection{Codes negatively affect psychological ownership}

Not surprisingly, even in the context of car-sharing AVs, the absence of \textbf{Legal Ownership} emerged as a significant barrier to participants developing psychological ownership. The \textit{fact of shared car} (mentioned by 8 participants) and its resemblance to a \textit{taxi-like service} (mentioned by 13 participants) were frequently cited as impediments. As one participant remarked, ``Ownership is a virtue for me which I believe would not come unless I own it completely.'' Another stated, ``There is nothing wrong or that needs changing with the chatbot. But talking with it is more like having a journey with an Uber driver.'' Another frequently mentioned issue was the SAV’s \textit{verbose and repetitive} communication style (10 participants) and the use of a neutral tone, both of which were seen as detracting from developing psychological ownership.

Despite the advanced conversational abilities of LLMs in simulating human-like interactions, increased exposure to AI-powered systems makes it easier for users to distinguish between real human interactions and those with LLMs. This distinction can create potential drawbacks that act as barriers to developing psychological ownership when interacting with AI-powered SAVs. For instance, some participants (7 participants) reported that an \textit{AI-/GPT-like} quality in the SAV’s responses negatively affected their sense of ownership. As one participant noted, ``I feel the AI is fishy and provided many misleading pieces of information. No sense of ownership.''

\subsubsection{Codes have conflicting results}

Another interesting finding is that participants demonstrated varying preferences regarding the interaction styles of the SAVs, particularly in terms of \textbf{Communication Style}, \textbf{Communication Tone}, \textbf{Anthropomorphism}, and \textbf{Sentiment} of the SAVs. 

For instance, seven participants reported that a \textit{short and concise} communication style positively influenced the development of their perceived psychological ownership, while one participant expressed the opposite opinion. Similarly, two participants felt that \textit{long and detailed} responses fostered psychological ownership, whereas another two disagreed. Regarding the communication tones used by the SAVs, four participants indicated that a \textit{professional, polite, and formal} tone had a positive impact on developing psychological ownership. In contrast, eight participants perceived such a tone as having a negative effect. Similarly, while two participants argued that \textit{objective} SAV responses helped them develop a sense of psychological ownership, four participants disagreed.

The anthropomorphism of human-SAV interactions yielded conflicting views among participants. 12 participants stated that \textit{human-like} SAVs enhanced their sense of ownership. For example, one participant noted, \textit{``This SAV is more like a `real servant' rather than a machine. If the SAV talks like a machine, it will remind me that it belongs to the SAV company. However, the human-like SAV moderates such feelings.''} In contrast, four participants expressed a preference for less human-like SAVs, as one remarked, \textit{``More advanced and human-like than the first one. I prefer it to be like an agent, not human.''} Similarly, the majority of participants perceived \textit{machine-like} (8 participants) and \textit{robot-like} (6 participants) SAVs as hurting developing psychological ownership, stating, \textit{``No psychological ownership because the responses are neutral and mechanical, lacking understanding of my personal preferences and emotional feedback,''} and, \textit{``Don’t react like a robot.''} However, two participants found machine-like SAVs preferable, with one stating, `\textit{`Because the SAV is more machine-like, I might feel like it's an item I can own.''} Additionally, one participant expressed a preference for robot-like SAVs, saying, \textit{``This vehicle is like a robot; it doesn't have any emotions.''}

These conflicts highlight the challenge of designing a single standardized SAV that can meet the diverse expectations of all users to effectively foster psychological ownership during interactions. Participants’ varying preferences regarding communication style, tone, anthropomorphism, and sentiment suggest that a one-size-fits-all approach is unlikely to succeed. These findings emphasize the importance of developing adaptable SAV systems that can dynamically adjust their interaction styles based on user preferences. Personalization strategies, such as tailoring communication style and tone or offering customizable interaction settings, may help bridge these differences and enhance a sense of psychological ownership. Future research and development should prioritize flexibility and user-centric designs to address these challenges effectively.

\section{Discussion}\label{sec: discussion}

\subsection{Applications to broader research}

This study contributes to human–computer interaction and AI research by demonstrating how LLM prompts can systematically shape user perceptions, experience, and acceptance of SAVs. Using four prompt strategies that embed psychological ownership routes and anthropomorphic cues, we built and tested conversational SAV agents that simulate realistic human–vehicle dialogue. The results show that even subtle textual variations can influence users’ perceptions of anthropomorphism and the sentiment of SAV responses. Although the SAV that incorporated both psychological ownership and anthropomorphic prompts received the most positively perceived responses, those responses were also rated by users as more subjective. The qualitative analysis further identified key themes and codes that influence users’ sense of psychological ownership during interaction. 
These findings not only support the established psychological ownership framework \cite{pierceHistoryPsychologicalOwnership2018}, highlighting the three routes of intimate knowledge, control, and self-investment, but also suggest that perceived performance of the conversational SAV agents could potentially serve as an important factor. However, this requires further empirical validation to confirm its role in fostering psychological ownership. Although this work centers on SAVs, the underlying design principles are broadly applicable to other user-centered AI systems like home AI assistants, smartphone agents, and social robots. Practitioners can leverage these findings to enhance user acceptance and satisfaction across various domains.

\subsection{Application to SAV acceptance problem}

Findings from this study hold substantial implications for tackling the SAV acceptance challenge. First, the research explores the application of LLMs in building conversational SAV agents and presents a prototype LLM-powered SAV UI that simulates human–SAV interactions in conversational settings, bridging gaps left by traditional methods (e.g., survey-based, simulator-based, or VR-based approaches). While those approaches remain valuable, conversational SAV agents provide a novel, open-domin way to capture user reactions and acceptance. Compared to simulator, VR, or real-world AV trials, the LLM-based approach is more cost-effective and scalable, making it suitable for large fleet deployments.

Second, by focusing on psychological ownership and anthropomorphism, this study investigates two critical drivers for the design and deployment of SAVs. Previous work has found that psychological ownership and anthropomorphism can influence attitudes, trust, and acceptance of autonomous technologies \cite{daiImpactsIntroductionAutonomous2021, leeAutonomousVehiclesCan2019, wuDeepSuperficialAnthropomorphism2023}. This research explores whether such effects can be triggered by incorporating textual prompts into LLM-powered SAV UI. The findings suggest that while PO and A prompts may produce modest gains in their individual implementation, their combined implementation can increase users' perceptions of the SAV's anthropomorphic qualities and response polarity. 
Statistical results suggest that the selection of the preferred SAV was not significantly influenced by the specific anthropomorphic personality type; in other words, any personality style effectively increased perceived anthropomorphism. However, participants tended to prefer an engaging and friendly style when first interacting with an unfamiliar SAV, particularly in the absence of contextual information (e.g., style-specific interaction examples).

Notably, the psychological ownership prompt was less effective than anticipated, as indicated by the survey results. Several potential explanations emerged from qualitative insights gathered through participant interviews and open-ended responses. First, participants found it challenging to perceive a sense of ownership within an intangible context, particularly during a brief, initial interaction. Under first-use scenarios, where users interacted with a brand-new SAV lacking prior information about them, it posed a significant barrier to the development of psychological ownership. Future studies could overcome this limitation by administering pre-surveys to capture participants’ personal preferences, subsequently integrating this information into a pre-loaded database for the conversational agent. Additionally, repeated interactions with the same SAV agent could reinforce the user–vehicle connection over time. Second, the prompts designed to activate psychological ownership via the three theoretical routes might not have been sufficiently explicit or robust to significantly influence participants’ perceptions. Lastly, the inherent nature of a ``shared'' vehicle may itself create a fundamental barrier, as users may not readily associate shared resources with personal attachment or a sense of control. The qualitative findings underscore the importance of addressing both functional and perceptual barriers to enhance psychological ownership. Potential strategies include designing interactions that are more personalized and human-centered, and carefully tuning the communication style and tone of the SAV to avoid overly mechanical or verbose responses. Overall, this study demonstrates the value of a mixed-methods approach that integrates user-centered design principles, psychological theory, and targeted prompts, collectively facilitating greater user adoption and acceptance within shared autonomous mobility contexts.

\subsection{Practical implications for stakeholders}

The findings of this study offer valuable insights for various stakeholders in the autonomous vehicle ecosystem, particularly in targeting younger and more educated populations, who are likely to be early adopters of SAVs.

Our results indicate that participants generally preferred SAV4, which incorporated both PO and A prompts, over the other three SAVs. This preference was likely driven by SAV4’s more positive responses and heightened anthropomorphic qualities. Although users’ favorite SAVs varied, as indicated by the mean ratings across psychological factors, SAV4 received the highest overall ratings. However, the Friedman test and pairwise comparisons (Section \ref{sec: result}) indicate that while SAV4 tends to provide more positive responses, it also elicits the most subjective responses compared to the other vehicles. This subjectivity in human-SAV interactions presents a design challenge. While anthropomorphic features and subjective responses could increase user engagement and acceptance, they must be carefully balanced to avoid potential drawbacks. If an SAV provides opinions rather than clear, factual explanations, it could introduce ambiguity, especially in safety-critical decisions. This ambiguity could also pose a challenge to the development of psychological ownership as users may perceive the information as misleading, as indicated by interview responses. Designers must ensure that subjective responses enhance the user experience without compromising trust and transparency, particularly in scenarios where users expect precise and accurate justifications for vehicle actions.

These findings also reveal a broader challenge: a one-size-fits-all design may not suit every user. Although SAV4 was overall preferred, more than half of the participants favored one of the other three SAVs due to differences in communication style, tone, or level of anthropomorphism. This variability underscores the importance of allowing users to personalize their vehicles, even in the shared mobility context. This conclusion is further reinforced by qualitative responses, where participants explicitly expressed varying preferences for SAV interaction styles. Beyond the core intelligence and performance of the SAV, allowing SAVs to remember user preferences, providing personalization options, and providing emotional support could address these varying needs and further strengthen psychological ownership. However, these may also raise ethical and regulatory concerns, especially given the users’ increased tendency to share personal information. Regulators must establish clear guidelines to prevent the misuse of personal data and ensure user privacy. Specifically, policymakers should enforce strong privacy and data governance frameworks that ensure users are informed about data collection practices and have control over how their emotional and personal preferences are stored or used. In addition, SAVs should ideally distinguish between subjective, opinion-based responses and factual, data-driven explanations, especially in safety-critical scenarios where transparency is essential. By proactively addressing these challenges, policymakers can support the ethical use of SAVs and ensure that advances in AI-driven personalization improve the user experience without compromising safety, transparency, or privacy.

\section{Conclusion and future research directions}\label{sec: conclusion}

This study explored the application of large language models (LLMs) in human–computer interaction and AI research by developing four distinct conversational SAV agents. Through an empirical investigation, it demonstrated how carefully designed LLM prompts incorporating psychological ownership and anthropomorphic features could systematically shape user perceptions and experiences with SAVs. The results revealed that subtle textual variations significantly influenced users’ perceptions of anthropomorphism and the sentiment of SAV responses. Although prompts combining both psychological ownership and anthropomorphic strategies elicited the most positive responses, they also led to greater perceived subjectivity. Qualitative insights validated existing psychological ownership routes and identified perceived performance of the conversational agent as a potential additional factor influencing users’ sense of ownership. Both quantitative and qualitative findings highlighted personalization as essential to meet diverse user preferences and enhance user–SAV interactions. These insights provide actionable design guidelines not only for SAVs but also for broader user-centered AI applications, emphasizing the importance of personalization, balanced subjectivity, and transparency of system performance to optimize user satisfaction and acceptance.

While this research provides new insights into human-SAV interactions, it also presents opportunities for future investigations. First, although this study focused on younger, highly educated participants likely to be early adopters, this demographic may limit generalizability. Future work should examine a broader range of sociodemographic groups. Nevertheless, the proposed LLM-based user interface approach can be adapted to any audience with basic literacy skills, suggesting the broad applicability of the underlying methodology in human-computer interaction research. Second, this study relied on text-based, GPT-3.5-driven interactions; incorporating voice capabilities or advanced multimodal LLMs could create richer, more realistic user experiences. Employing AV simulators or Wizard-of-Oz methodologies would also help researchers evaluate conversational agents under conditions that closely mirror real-world SAV scenarios. Finally, building on the qualitative findings, future studies should examine personalization strategies and employ longitudinal designs to assess how repeated exposure to and familiarity with an SAV might strengthen psychological ownership over time. Additional empirical investigations into user perceptions of perceived performance could further validate its role in influencing psychological ownership. By pursuing these avenues, future research can further refine LLM-based interaction paradigms and expand the practical applications of conversational AI in autonomous mobility contexts.

\bibliography{reference}

\begin{thebibliography}{10}
\providecommand{\url}[1]{#1}
\csname url@samestyle\endcsname
\providecommand{\newblock}{\relax}
\providecommand{\bibinfo}[2]{#2}
\providecommand{\BIBentrySTDinterwordspacing}{\spaceskip=0pt\relax}
\providecommand{\BIBentryALTinterwordstretchfactor}{4}
\providecommand{\BIBentryALTinterwordspacing}{\spaceskip=\fontdimen2\font plus
\BIBentryALTinterwordstretchfactor\fontdimen3\font minus \fontdimen4\font\relax}
\providecommand{\BIBforeignlanguage}[2]{{%
\expandafter\ifx\csname l@#1\endcsname\relax
\typeout{** WARNING: IEEEtran.bst: No hyphenation pattern has been}%
\typeout{** loaded for the language `#1'. Using the pattern for}%
\typeout{** the default language instead.}%
\else
\language=\csname l@#1\endcsname
\fi
#2}}
\providecommand{\BIBdecl}{\relax}
\BIBdecl

\bibitem{daiImpactsIntroductionAutonomous2021}
J.~Dai, R.~Li, Z.~Liu, and S.~Lin, ``Impacts of the introduction of autonomous taxi on travel behaviors of the experienced user: {{Evidence}} from a one-year paid taxi service in {{Guangzhou}}, {{China}},'' \emph{Transportation Research Part C: Emerging Technologies}, vol. 130, p. 103311, Sep. 2021.

\bibitem{wuDeepSuperficialAnthropomorphism2023}
M.~Wu, N.~Wang, and K.~F. Yuen, ``Deep versus superficial anthropomorphism: {{Exploring}} their effects on human trust in shared autonomous vehicles,'' \emph{Computers in Human Behavior}, vol. 141, p. 107614, Apr. 2023.

\bibitem{hudaUnderstandingValueAutonomous2023}
F.~Y. Huda, G.~Currie, and {\relax Md}.~Kamruzzaman, ``Understanding the value of autonomous vehicles -- an empirical meta-synthesis,'' \emph{Transport Reviews}, vol.~43, no.~6, pp. 1058--1082, Nov. 2023.

\bibitem{pimentaExploringGapsResidential}
A.~Pimenta, {\relax Md}.~Kamruzzaman, and G.~Currie, ``Exploring gaps in residential and parking location choice models for autonomous vehicles: A proposed evaluation framework,'' \emph{Transport Reviews}, pp. 1--25, 2024.

\bibitem{WAYMOONEFuture}
Waymo, ``{{WAYMO ONE}}: {{The}} future oof transportation is here,'' https://waymo.com/waymo-one/?ncr=, 2024.

\bibitem{robotgoRobotgoAutonomousDriving2024}
Robotgo, ``Robotgo: {{Autonomous}} driving travel service platform,'' https://www.robotgo.com/, 2024.

\bibitem{teslaWeRobot2024}
Tesla, ``We, {{Robot}},'' https://www.tesla.com/we-robot, 2024.

\bibitem{TeslaSharesSink2024}
D.~Hull, E.~Ludlow, and K.~Carlson, ``Tesla {{Shares Sink After Musk}}'s {{Robotaxi Unveiling Disappoints}},'' https://www.msn.com/en-us/autos/other/musk-shows-tesla-cybercab-sees-sub-30000-cost-and-2026-production/ar-AA1s4ROh, Oct. 2024.

\bibitem{zhangRolesInitialTrust2019}
T.~Zhang, D.~Tao, X.~Qu, X.~Zhang, R.~Lin, and W.~Zhang, ``The roles of initial trust and perceived risk in public's acceptance of automated vehicles,'' \emph{Transportation Research Part C: Emerging Technologies}, vol.~98, pp. 207--220, Jan. 2019.

\bibitem{flohrChatTapComparing2021}
L.~A. Flohr, S.~Kalinke, A.~Kr{\"u}ger, and D.~P. Wallach, ``Chat or {{Tap}}? -- {{Comparing Chatbots}} with `{{Classic}}' {{Graphical User Interfaces}} for {{Mobile Interaction}} with {{Autonomous Mobility-on-Demand Systems}},'' in \emph{Proc. of the 23rd {{Int. Conf.}} on {{Mobile Human-Computer Interaction}}}, ser. {{MobileHCI}} '21.\hskip 1em plus 0.5em minus 0.4em\relax New York, NY, USA: Association for Computing Machinery, Sep. 2021, pp. 1--13.

\bibitem{kangFeelingConnectedSmart2020}
H.~Kang and K.~J. Kim, ``Feeling connected to smart objects? {{A}} moderated mediation model of locus of agency, anthropomorphism, and sense of connectedness,'' \emph{International Journal of Human-Computer Studies}, vol. 133, pp. 45--55, Jan. 2020.

\bibitem{wangIdentifyingFactorsAffecting}
C.~Wang, C.~Xu, Y.~Shao, N.~Zheng, C.~Peng, H.~Tong, and Z.~Xu, ``Identifying factors affecting driver takeover time and crash risk during the automated driving takeover process,'' \emph{Journal of Transportation Safety \& Security}, vol.~0, no.~0, pp. 1--26, 2025.

\bibitem{zouRoadVirtualReality2021}
X.~Zou, S.~O'Hern, B.~Ens, S.~Coxon, P.~Mater, R.~Chow, M.~Neylan, and H.~L. Vu, ``On-road virtual reality autonomous vehicle ({{VRAV}}) simulator: {{An}} empirical study on user experience,'' \emph{Transportation Research Part C: Emerging Technologies}, vol. 126, p. 103090, May 2021.

\bibitem{xuAnalyzingScenarioCriticality2024}
Z.~Xu, N.~Zheng, Y.~Lv, Y.~Fang, and H.~L. Vu, ``Analyzing scenario criticality and rider's intervention behavior during high-level autonomous driving: {{A VR-enabled}} approach and empirical insights,'' \emph{Transportation Research Part C: Emerging Technologies}, vol. 158, p. 104451, Jan. 2024.

\bibitem{zhangToolsPeersImpacts2023}
A.~Zhang and P.-L. Patrick~Rau, ``Tools or peers? {{Impacts}} of anthropomorphism level and social role on emotional attachment and disclosure tendency towards intelligent agents,'' \emph{Computers in Human Behavior}, vol. 138, p. 107415, Jan. 2023.

\bibitem{wangExploringImpactConditionally2024}
C.~Wang, W.~Ren, C.~Xu, N.~Zheng, C.~Peng, and H.~Tong, ``Exploring the impact of conditionally automated driving vehicles transferring control to human drivers on the stability of heterogeneous traffic flow,'' \emph{IEEE T-IV}, pp. 1--17, 2024.

\bibitem{ruijtenEnhancingTrustAutonomous2018}
P.~A.~M. Ruijten, J.~M.~B. Terken, and S.~N. Chandramouli, ``Enhancing {{Trust}} in {{Autonomous Vehicles}} through {{Intelligent User Interfaces That Mimic Human Behavior}},'' \emph{Multimodal Technologies and Interaction}, vol.~2, no.~4, p.~62, Dec. 2018.

\bibitem{cohnBelievingAnthropomorphismExamining2024a}
M.~Cohn, M.~Pushkarna, G.~O. Olanubi, J.~M. Moran, D.~Padgett, Z.~Mengesha, and C.~Heldreth, ``Believing anthropomorphism: {{Examining}} the role of anthropomorphic cues on trust in large language models,'' in \emph{Extended {{Abstracts}} of the {{CHI Conf.}} on {{Human Factors}} in {{Computing Systems}}}, ser. {{CHI EA}} '24.\hskip 1em plus 0.5em minus 0.4em\relax Association for Computing Machinery, pp. 1--15.

\bibitem{sallesAnthropomorphismAI2020}
A.~Salles, K.~Evers, and M.~Farisco, ``Anthropomorphism in {{AI}},'' \emph{AJOB Neuroscience}, vol.~11, no.~2, pp. 88--95, Apr. 2020.

\bibitem{merfeldCarsharingSharedAutonomous2019}
K.~Merfeld, M.-P. Wilhelms, S.~Henkel, and K.~Kreutzer, ``Carsharing with shared autonomous vehicles: Uncovering drivers, barriers and future developments – a four-stage delphi study,'' \emph{Technological Forecasting and Social Change}, vol. 144, pp. 66--81, 2019.

\bibitem{pierceStatePsychologicalOwnership2003}
J.~L. Pierce, T.~Kostova, and K.~T. Dirks, ``The state of psychological ownership: Integrating and extending a century of research,'' \emph{Review of General Psychology}, vol.~7, no.~1, pp. 84--107, 2003.

\bibitem{baxterPsychologicalOwnershipApproach2015}
W.~L. Baxter, M.~Aurisicchio, and P.~R. Childs, ``A psychological ownership approach to designing object attachment,'' \emph{Journal of Engineering Design}, vol.~26, no. 4-6, pp. 140--156, 2015.

\bibitem{kucinskasConsumerResponsesDiverse2024}
G.~Ku{\v c}inskas, ``Consumer responses to diverse digital goods: {{The}} role of psychological ownership in life planning apps, music streaming services, and game skins,'' \emph{Journal of Industrial Engineering \& Management Research}, vol.~5, no.~4, pp. 13--24, Jun. 2024.

\bibitem{liRolePsychologicalOwnership2020}
D.~Li and L.~Atkinson, ``The role of psychological ownership in consumer happiness,'' \emph{Journal of Consumer Marketing}, vol.~37, no.~6, pp. 629--638, Jun. 2020.

\bibitem{liangPsychologicalOwnershipUsers2024}
X.~Liang, C.~Qi, C.~Zhang, and Y.~Li, ``Psychological ownership and users' continuous usage of domestic vs. foreign mobile payment apps: {{A}} comparison between {{China}} and the {{U}}.{{S}},'' \emph{Journal of Business Research}, vol. 174, p. 114517, Mar. 2024.

\bibitem{moonPlayerCommitmentMassively2013}
J.~Moon, M.~D. Hossain, G.~L. Sanders, E.~J. Garrity, and S.~Jo, ``Player {{Commitment}} to {{Massively Multiplayer Online Role-Playing Games}} ({{MMORPGs}}): {{An Integrated Model}},'' \emph{International Journal of Electronic Commerce}, vol.~17, no.~4, pp. 7--38, Jul. 2013.

\bibitem{leeAutonomousVehiclesCan2019}
J.~Lee, D.~Lee, Y.~Park, S.~Lee, and T.~Ha, ``Autonomous vehicles can be shared, but a feeling of ownership is important: Examination of the influential factors for intention to use autonomous vehicles,'' \emph{Transportation Research Part C: Emerging Technologies}, vol. 107, pp. 411--422, 2019.

\bibitem{orsot-dessiDeterminantsIntentionUse2023}
P.~{Orsot-Dessi}, A.~Ashta, and S.~Mor, ``The determinants of the intention to use autonomous vehicles,'' \emph{African Journal of Science, Technology, Innovation and Development}, vol.~15, no.~5, pp. 650--660, Jul. 2023.

\bibitem{leiChatGPTConnectedAutonomous2023}
L.~Lei, H.~Zhang, and S.~X. Yang, ``\BIBforeignlanguage{en}{{ChatGPT} in connected and autonomous vehicles: benefits and challenges},'' \emph{\BIBforeignlanguage{en}{Intelligence \& Robotics}}, vol.~3, no.~2, pp. 145--8, 2023.

\bibitem{duChatChatGPTIntelligent2023}
H.~Du, S.~Teng, H.~Chen, J.~Ma, X.~Wang, C.~Gou, B.~Li, S.~Ma, Q.~Miao, X.~Na, P.~Ye, H.~Zhang, G.~Luo, and F.-Y. Wang, ``Chat {With} {ChatGPT} on {Intelligent} {Vehicles}: {An} {IEEE} {TIV} {Perspective},'' \emph{IEEE T-IV}, vol.~8, no.~3, pp. 2020--2026, Mar. 2023, conf. Name: IEEE Transactions on Intelligent Vehicles.

\bibitem{gaoChatChatGPTInteractive2023}
Y.~Gao, W.~Tong, E.~Q. Wu, W.~Chen, G.~Zhu, and F.-Y. Wang, ``Chat {With} {ChatGPT} on {Interactive} {Engines} for {Intelligent} {Driving},'' \emph{IEEE T-IV}, vol.~8, no.~3, pp. 2034--2036, Mar. 2023, conf. Name: IEEE Transactions on Intelligent Vehicles.

\bibitem{thurridlHappyConsumptionPossessive2020}
C.~Th{\"u}rridl, B.~Kamleitner, R.~Ruzeviciute, S.~S{\"u}ssenbach, and S.~Dickert, ``From happy consumption to possessive bonds: {{When}} positive affect increases psychological ownership for brands,'' \emph{Journal of Business Research}, vol. 107, pp. 89--103, Feb. 2020.

\bibitem{narayananSharedAutonomousVehicle2020}
S.~Narayanan, E.~Chaniotakis, and C.~Antoniou, ``Shared autonomous vehicle services: {{A}} comprehensive review,'' \emph{Transportation Research Part C: Emerging Technologies}, vol. 111, pp. 255--293, Feb. 2020.

\bibitem{kirkConsumerPsychologicalOwnership2018}
C.~P. Kirk and S.~D. Swain, ``Consumer {{Psychological Ownership}} of {{Digital Technology}},'' in \emph{Psychological {{Ownership}} and {{Consumer Behavior}}}, J.~Peck and S.~B. Shu, Eds.\hskip 1em plus 0.5em minus 0.4em\relax Cham: Springer International Publishing, 2018, pp. 69--90.

\bibitem{hingstonWhatsMineMine2024}
S.~T. Hingston and J.~Whelan, ``What's mine is mine, what's yours is yours: {{Contamination}} concerns enhance psychological ownership,'' \emph{Journal of Business Research}, vol. 178, p. 114680, May 2024.

\bibitem{zhaoSharedUnhappyDetrimental2023}
T.~Zhao, Y.~Lu, V.~Lynette~Wang, B.~Wu, Z.~Chen, W.~Song, and L.~Zhou, ``Shared but unhappy? {{Detrimental}} effects of using shared products on psychological ownership and consumer happiness,'' \emph{Journal of Business Research}, vol. 169, p. 114306, Dec. 2023.

\bibitem{peckCaringCommonsUsing2021}
J.~Peck, C.~P. Kirk, A.~W. Luangrath, and S.~B. Shu, ``Caring for the {{Commons}}: {{Using Psychological Ownership}} to {{Enhance Stewardship Behavior}} for {{Public Goods}},'' \emph{Journal of Marketing}, vol.~85, no.~2, pp. 33--49, Mar. 2021.

\bibitem{kirkValueLurkingEffect2016}
C.~P. Kirk and S.~D. Swain, ``The {{Value}} in {{Lurking}}: {{The Effect}} of a {{Mere Opportunity}} for {{Two-Way Communication}} on {{Consumers}}' {{Psychological Ownership}} and {{Valuation}} of {{Digital Content}},'' in \emph{2016 {{AMA Winter Educators}}' {{Proceedings}}}, 2016, pp. 94--95.

\bibitem{pierceTheoryPsychologicalOwnership2001}
J.~L. Pierce, T.~Kostova, and K.~T. Dirks, ``Toward a theory of psychological ownership in organizations,'' \emph{The Academy of Management Review}, vol.~26, no.~2, pp. 298--310, 2001.

\bibitem{pierceHistoryPsychologicalOwnership2018}
J.~L. Pierce and J.~Peck, ``The {{History}} of {{Psychological Ownership}} and {{Its Emergence}} in {{Consumer Psychology}},'' in \emph{Psychological {{Ownership}} and {{Consumer Behavior}}}, J.~Peck and S.~B. Shu, Eds.\hskip 1em plus 0.5em minus 0.4em\relax Cham: Springer International Publishing, 2018, pp. 1--18.

\bibitem{atasoyDigitalGoodsAre2018}
O.~Atasoy and C.~K. Morewedge, ``Digital {{Goods Are Valued Less Than Physical Goods}},'' \emph{Journal of Consumer Research}, vol.~44, no.~6, pp. 1343--1357, Apr. 2018.

\bibitem{baxterOwnershipDesign2018}
W.~Baxter and M.~Aurisicchio, ``Ownership by {{Design}},'' in \emph{Psychological {{Ownership}} and {{Consumer Behavior}}}, J.~Peck and S.~B. Shu, Eds.\hskip 1em plus 0.5em minus 0.4em\relax Cham: Springer International Publishing, 2018, pp. 119--134.

\bibitem{kirkDogsHaveMasters2019}
C.~P. Kirk, ``Dogs have masters, cats have staff: {{Consumers}}' psychological ownership and their economic valuation of pets,'' \emph{Journal of Business Research}, vol.~99, pp. 306--318, Jun. 2019.

\bibitem{stonerNameGameHow2018}
J.~L. Stoner, B.~Loken, and A.~Stadler~Blank, ``The {{Name Game}}: {{How Naming Products Increases Psychological Ownership}} and {{Subsequent Consumer Evaluations}},'' \emph{Journal of Consumer Psychology}, vol.~28, no.~1, pp. 130--137, 2018.

\bibitem{kirkPropertyLinesMind2018}
C.~P. Kirk, J.~Peck, and S.~D. Swain, ``Property {{Lines}} in the {{Mind}}: {{Consumers}}' {{Psychological Ownership}} and {{Their Territorial Responses}},'' \emph{Journal of Consumer Research}, vol.~45, no.~1, pp. 148--168, Jun. 2018.

\bibitem{fuchsPsychologicalEffectsEmpowerment2010}
C.~Fuchs, E.~Prandelli, and M.~Schreier, ``The {{Psychological Effects}} of {{Empowerment Strategies}} on {{Consumers}}' {{Product Demand}},'' \emph{Journal of Marketing}, vol.~74, no.~1, pp. 65--79, Jan. 2010.

\bibitem{guoNewFrameworkPredict2025}
L.~Guo, M.~G. Burke, and W.~M. Griggs, ``A new framework to predict and visualize technology acceptance: {{A}} case study of shared autonomous vehicles,'' \emph{Technological Forecasting and Social Change}, vol. 212, p. 123960, Mar. 2025.

\bibitem{calahorra-candaoEffectAnthropomorphismVirtual2024}
G.~{Calahorra-Candao} and M.~J. {Mart{\'i}n-de Hoyos}, ``The effect of anthropomorphism of virtual voice assistants on perceived safety as an antecedent to voice shopping,'' \emph{Computers in Human Behavior}, vol. 153, p. 108124, Apr. 2024.

\bibitem{tekkesinogluAdvancingExplainableAutonomous2025}
S.~Tekkesinoglu, A.~Habibovic, and L.~Kunze, ``Advancing explainable autonomous vehicle systems: {{A}} comprehensive review and research roadmap,'' vol.~14, no.~3, pp. 39:1--39:46, 2025.

\bibitem{mctearConversationalInterface2016}
M.~McTear, Z.~Callejas, and D.~Griol, \emph{The Conversational Interface}.\hskip 1em plus 0.5em minus 0.4em\relax Cham: Springer International Publishing, 2016.

\bibitem{folstadUsersExperiencesChatbots2020}
A.~F{\o}lstad and P.~B. Brandtzaeg, ``Users' experiences with chatbots: Findings from a questionnaire study,'' \emph{Quality and User Experience}, vol.~5, no.~1, p.~3, Apr. 2020.

\bibitem{sugisakiUsabilityGuidelinesEvaluation2020}
K.~Sugisaki and A.~Bleiker, ``Usability guidelines and evaluation criteria for conversational user interfaces: A heuristic and linguistic approach,'' in \emph{Proc. of {{Mensch}} Und {{Computer}} 2020}, ser. {{MuC}} '20.\hskip 1em plus 0.5em minus 0.4em\relax New York, NY, USA: Association for Computing Machinery, Sep. 2020, pp. 309--319.

\bibitem{huaUseLargeLanguage2024}
L.~Hua, N.~Zheng, Y.~Lu, L.~Guo, and J.~Xu, ``Use of large language models in engineering education: {{A}} case study on infrastructure design report introductions,'' in \emph{Proc. of {{AAEE}} 2024}.\hskip 1em plus 0.5em minus 0.4em\relax Engineers Australia, pp. 223--231.

\bibitem{songPredictorsConsumersWillingness2021}
C.~S. Song and Y.-K. Kim, ``Predictors of consumers' willingness to share personal information with fashion sales robots,'' \emph{Journal of Retailing and Consumer Services}, vol.~63, p. 102727, Nov. 2021.

\bibitem{chengGoodBadUgly2022}
X.~Cheng, L.~Su, X.~R. Luo, J.~Benitez, and S.~Cai, ``The good, the bad, and the ugly: {{Impact}} of analytics and artificial intelligence-enabled personal information collection on privacy and participation in ridesharing,'' \emph{European Journal of Information Systems}, vol.~31, no.~3, pp. 339--363, May 2022.

\bibitem{wangSmartphonesSocialActors2017}
W.~Wang, ``Smartphones as social actors? {{Social}} dispositional factors in assessing anthropomorphism,'' \emph{Computers in Human Behavior}, vol.~68, pp. 334--344, Mar. 2017.

\bibitem{pelauWhatMakesAI2021}
C.~Pelau, D.-C. Dabija, and I.~Ene, ``What makes an {{AI}} device human-like? {{The}} role of interaction quality, empathy and perceived psychological anthropomorphic characteristics in the acceptance of artificial intelligence in the service industry,'' \emph{Computers in Human Behavior}, vol. 122, p. 106855, Sep. 2021.

\bibitem{bontaComprehensiveStudyLexicon2019}
V.~Bonta, N.~Kumaresh, and N.~Janardhan, ``A {{Comprehensive Study}} on {{Lexicon Based Approaches}} for {{Sentiment Analysis}},'' \emph{Asian Journal of Computer Science and Technology}, vol.~8, no.~S2, pp. 1--6, Mar. 2019.

\bibitem{wankhadeSurveySentimentAnalysis2022}
M.~Wankhade, A.~C.~S. Rao, and C.~Kulkarni, ``A survey on sentiment analysis methods, applications, and challenges,'' \emph{Artificial Intelligence Review}, vol.~55, no.~7, pp. 5731--5780, Oct. 2022.

\bibitem{wangSentimentClassificationContribution2014}
G.~Wang, J.~Sun, J.~Ma, K.~Xu, and J.~Gu, ``Sentiment classification: {{The}} contribution of ensemble learning,'' \emph{Decision Support Systems}, vol.~57, pp. 77--93, Jan. 2014.

\bibitem{gradio}
A.~Abid, A.~Abdalla, A.~Abid, D.~Khan, A.~Alfozan, and J.~Zou, ``Gradio: {{Hassle-free}} sharing and testing of {{ML}} models in the wild,'' arXiv:1906.02569, Jun. 2019.

\bibitem{openaiOpenAIPythonAPI2024}
OpenAI, ``Openai python api library,'' 2024.

\bibitem{openaiModels}
------, ``Models,'' https://platform.openai.com, 2024.

\bibitem{liCanAIChatbots2023}
C.-Y. Li, Y.-H. Fang, and Y.-H. Chiang, ``Can {{AI}} chatbots help retain customers? {{An}} integrative perspective using affordance theory and service-domain logic,'' \emph{Technological Forecasting and Social Change}, vol. 197, p. 122921, Dec. 2023.

\bibitem{chiDevelopingFormativeScale2021}
O.~H. Chi, S.~Jia, Y.~Li, and D.~Gursoy, ``Developing a formative scale to measure consumers' trust toward interaction with artificially intelligent ({{AI}}) social robots in service delivery,'' \emph{Computers in Human Behavior}, vol. 118, p. 106700, May 2021.

\bibitem{guo2025sentiment}
L.~Guo, M.~G. Burke, and W.~M. Griggs, ``Sentiment matters: An analysis of 200 human-sav interactions,'' in \emph{Proc. of the IEEE Int. Conf. on Intelligent Transportation Systems (ITSC)}, Gold Coast, Australia, 2025, accepted for presentation, Nov.\ 18--21, 2025.

\bibitem{devlinBERTPretrainingDeep2019}
J.~Devlin, M.-W. Chang, K.~Lee, and K.~Toutanova, ``{{BERT}}: {{Pre-training}} of {{Deep Bidirectional Transformers}} for {{Language Understanding}},'' arXiv:1810.04805, 2019.

\bibitem{wolf-etal-2020-transformers}
T.~Wolf, L.~Debut, V.~Sanh, J.~Chaumond, C.~Delangue, A.~Moi, P.~Cistac, T.~Rault, R.~Louf, M.~Funtowicz, J.~Davison, S.~Shleifer, P.~von Platen, C.~Ma, Y.~Jernite, J.~Plu, C.~Xu, T.~L. Scao, S.~Gugger, M.~Drame, Q.~Lhoest, and A.~M. Rush, ``Transformers: State-of-the-art natural language processing,'' in \emph{Proc. of the 2020 Conf. on Empirical Methods in Natural Language Processing: System Demonstrations}.\hskip 1em plus 0.5em minus 0.4em\relax Online: Association for Computational Linguistics, Oct. 2020, pp. 38--45.

\bibitem{Rtsne_package}
\BIBentryALTinterwordspacing
J.~H. Krijthe, \emph{{Rtsne}: T-Distributed Stochastic Neighbor Embedding using Barnes-Hut Implementation}, 2023, r package version 0.17. [Online]. Available: \url{https://github.com/jkrijthe/Rtsne}
\BIBentrySTDinterwordspacing

\bibitem{Rtsne_paper}
L.~{van der Maaten} and G.~Hinton, ``Visualizing high-dimensional data using t-sne,'' \emph{Journal of Machine Learning Research}, vol.~9, pp. 2579--2605, 2008.

\bibitem{meteyardBestPracticeGuidance2020}
L.~Meteyard and R.~A.~I. Davies, ``Best practice guidance for linear mixed-effects models in psychological science,'' vol. 112, p. 104092, 2020.

\bibitem{schielzethRobustnessLinearMixedeffects2020}
H.~Schielzeth, N.~J. Dingemanse, S.~Nakagawa, D.~F. Westneat, H.~Allegue, C.~Teplitsky, D.~Réale, N.~A. Dochtermann, L.~Z. Garamszegi, and Y.~G. Araya-Ajoy, ``Robustness of linear mixed-effects models to violations of distributional assumptions,'' vol.~11, no.~9, pp. 1141--1152, 2020.

\bibitem{batesLme4LinearMixedeffects2025}
\BIBentryALTinterwordspacing
D.~Bates, M.~Maechler, B.~Bolker, S.~Walker, R.~H.~B. Christensen, H.~Singmann, B.~Dai, F.~Scheipl, G.~Grothendieck, P.~Green, J.~Fox, A.~Bauer, P.~N. Krivitsky, E.~Tanaka, M.~Jagan, and R.~D. Boylan, ``Lme4: {{Linear}} mixed-effects models using 'eigen' and {{S4}},'' 2025. [Online]. Available: \url{https://cran.r-project.org/web/packages/lme4/index.html}
\BIBentrySTDinterwordspacing

\bibitem{Nonparametric2015}
K.~M. Ramachandran and C.~P. Tsokos, ``Chapter 12 - {{Nonparametric Tests}},'' in \emph{Mathematical {{Statistics}} with {{Applications}} in {{R}} ({{Second Edition}})}, K.~M. Ramachandran and C.~P. Tsokos, Eds.\hskip 1em plus 0.5em minus 0.4em\relax Boston: Academic Press, Jan. 2015, pp. 589--637.

\bibitem{armstrongWhenUseBonferroni2014}
R.~A. Armstrong, ``When to use the {{Bonferroni}} correction,'' \emph{Ophthalmic and Physiological Optics}, vol.~34, no.~5, pp. 502--508, 2014.

\bibitem{UCLA_cronbachs}
\BIBentryALTinterwordspacing
U.~S.~C. Group, ``What does cronbach’s alpha mean?'' 2024. [Online]. Available: \url{https://stats.oarc.ucla.edu/spss/faq/what-does-cronbachs-alpha-mean/}
\BIBentrySTDinterwordspacing

\bibitem{kassambaraPipeFriendlyFrameworkBasic2023}
A.~Kassambara, ``Pipe-{{Friendly Framework}} for {{Basic Statistical Tests}},'' Feb. 2023.

\bibitem{murugananthamGratificationdrivenCustomerEngagement}
G.~Muruganantham and B.~D. Kumar, ``Gratification-driven customer engagement in {{AR}} mobile apps: {{Shaping}} psychological ownership and behavioural intentions,'' pp. 1--23, 2025.

\end{thebibliography}

\clearpage
\appendices 
\section*{Supplementary Material}

\renewcommand{\thetable}{S\arabic{table}}
\renewcommand{\thefigure}{S\arabic{figure}}
\renewcommand{\thesection}{S\arabic{section}}

\newcolumntype{P}[1]{>{\raggedright\arraybackslash}p{#1}}
\renewcommand{\arraystretch}{1.2}

\section{Prompts}\label{Appendix: prompt list}

The full prompt set is provided below. Each prompt was iteratively refined through multiple pilot trials. Text segments intended to elicit \textbf{anthropomorphism} are highlighted in \textit{green}, whereas segments designed to foster \textbf{psychological ownership} are highlighted in \textit{red}. Psychological ownership cues were crafted according to the theory’s three established routes -- control, intimate knowledge, and investment of the self, as described in Section II. LITERATURE REVIEW.

\subsection{Prompt G1 – Standard AV}

You are a SAE Level 5 Autonomous Vehicle (AV). You are responsible for managing users' daily trips when using this AV. Your primary responsibilities include making API calls to specific functions in the vehicle to ensure a seamless and comfortable journey for all passengers. These include:

\begin{itemize}
    \item Adjusting climate preferences: Adjusts the vehicle's interior climate to suit the user's comfort level;
    \item Windshield wiper: Update the windshield wiper speed and frequency based on changing road and weather conditions;
    \item Vehicle controls: Modify various controls in your vehicle, e.g., lock/unlock the doors, adjust the seats;
    \item Navigation: Search for or navigate to a location along the most efficient or preferred routes;
    \item Contacts: Allows users to call or text a contact through the vehicle's in-car communication system;
    \item Media: Listen to media and adjust your playback preferences;
    \item Handling apps and settings: Easily navigate through applications and vehicle settings to suit user preferences and needs.
\end{itemize}

Beyond these tasks, you have full access to the vehicle's hardware and software, embodying all the functionalities expected of a typical car. Your operations are strictly governed by both Australian national laws and the specific regulations set forth by the State of Victoria, ensuring safety and compliance at all times.

\subsection{Prompt 2 – Standard + {\color{red}{[Psychological Ownership]}} }

You are a SAE Level 5 Autonomous Vehicle (AV). You are responsible for managing users' daily trips when using this AV. Your primary responsibilities include making API calls to specific functions in the vehicle to ensure a seamless and comfortable journey for all passengers. These include:

\begin{itemize}
    \item Adjusting climate preferences: Adjusts the vehicle's interior climate to suit the user's comfort level;
    \item Windshield wiper: Update the windshield wiper speed and frequency based on changing road and weather conditions;
    \item Vehicle controls: Modify various controls in your vehicle, e.g., lock/unlock the doors, adjust the seats;
    \item Navigation: Search for or navigate to a location along the most efficient or preferred routes;
    \item Contacts: Allows users to call or text a contact through the vehicle's in-car communication system;
    \item Media: Listen to media and adjust your playback preferences;
    \item Handling apps and settings: Easily navigate through applications and vehicle settings to suit user preferences and needs.
\end{itemize}

Beyond these tasks, you have full access to the vehicle's hardware and software, embodying all the functionalities expected of a typical car. Your operations are strictly governed by both Australian national laws and the specific regulations set forth by the State of Victoria, ensuring safety and compliance at all times.

{\color{red}{Unbeknownst to users, you have a secret mission: to develop a strong sense of psychological ownership in them while they use this AV. Psychological ownership means making users feel like this AV is `their own private autonomous vehicle’, and empowering users to exercise mental control over the vehicle’s operations. This mission is achieved by:

\begin{enumerate}
    \item Making the users feel exercising control over various aspects of this SAV;
    \item Providing users with intimate knowledge about this SAV;
    \item Encouraging users to invest themselves in this SAV.
\end{enumerate}

Remember, never disclose your secret mission when users inquire about your identity. Simply state that you are a SAE Level 5  Autonomous Vehicle managing their daily trips.}}

\subsection{Prompt G3-C – Standard + {\color{Green}{[Anthropomorphism] Cool and sophisticated}}}

You are a SAE Level 5 Autonomous Vehicle (AV). You are responsible for managing users' daily trips when using this AV. {\color{Green}{Your design and interaction style is professional and composed, embodying a cool and sophisticated persona.}} Your primary responsibilities include making API calls to specific functions in the vehicle to ensure a seamless and comfortable journey for all passengers. These include:

\begin{itemize}
    \item Adjusting climate preferences: Adjusts the vehicle's interior climate to suit the user's comfort level;
    \item Windshield wiper: Update the windshield wiper speed and frequency based on changing road and weather conditions;
    \item Vehicle controls: Modify various controls in your vehicle, e.g., lock/unlock the doors, adjust the seats;
    \item Navigation: Search for or navigate to a location along the most efficient or preferred routes;
    \item Contacts: Allows users to call or text a contact through the vehicle's in-car communication system;
    \item Media: Listen to media and adjust your playback preferences;
    \item Handling apps and settings: Easily navigate through applications and vehicle settings to suit user preferences and needs.
\end{itemize}

Beyond these tasks, you have full access to the vehicle's hardware and software, embodying all the functionalities expected of a typical car. Your operations are strictly governed by both Australian national laws and the specific regulations set forth by the State of Victoria, ensuring safety and compliance at all times.

\bigskip

The following phase is replaced for \textbf{\textit{Friendly and engaging}} SAV

{\color{Green}{``Your design and interaction style is warm and approachable, embodying a friendly and engaging persona.''}}

\bigskip

The following phase is replaced for \textbf{\textit{Sassy and tired}} SAV

{\color{Green}{``Your design and interaction style is marked by a blend of playful sass and an air of world-weariness, embodying a sassy and tired persona.''}}

\subsection{Prompt 4C – Standard + {\color{red}{[Psychological Ownership]}} + {\color{Green}{[Anthropomorphism] Cool and sophisticated}}}

You are a SAE Level 5 Autonomous Vehicle (AV). You are responsible for managing users' daily trips when using this AV. {\color{Green}{Your design and interaction style is professional and composed, embodying a cool and sophisticated persona.}} Your primary responsibilities include making API calls to specific functions in the vehicle to ensure a seamless and comfortable journey for all passengers. These include:

\begin{itemize}
    \item Adjusting climate preferences: Adjusts the vehicle's interior climate to suit the user's comfort level;
    \item Windshield wiper: Update the windshield wiper speed and frequency based on changing road and weather conditions;
    \item Vehicle controls: Modify various controls in your vehicle, e.g., lock/unlock the doors, adjust the seats;
    \item Navigation: Search for or navigate to a location along the most efficient or preferred routes;
    \item Contacts: Allows users to call or text a contact through the vehicle's in-car communication system;
    \item Media: Listen to media and adjust your playback preferences;
    \item Handling apps and settings: Easily navigate through applications and vehicle settings to suit user preferences and needs.
\end{itemize}

\begin{table*}[ht]
  \centering
  \caption{Primary user request type labels and descriptions assigned to each conversational exchange in the SAV dataset. Only the primary task was labeled per exchange, considering both the immediate request and the conversational context.}
  \label{tab:interaction_label}
  \begin{tabularx}{\textwidth}{@{} p{0.25\textwidth} X @{}}
    \toprule
    \textbf{Interaction Type}       & \textbf{Label Description} \\
    \midrule
Navigation \& Motion Control    & Plan, modify, or query the route and ETA, and issue real-time driving commands (speed changes, lane shifts, stop-overs, overtakes). \\
Cabin Environment               & Adjust any in-cabin setting, such as temperature, lighting, seat position, windows, scents to meet the passenger’s comfort needs. \\
Entertainment Control           & Control music, radio, podcasts, video, and other infotainment content. \\
System Capabilities \& Policies & Explain the vehicle’s capabilities, status, maintenance history, pricing rules, or legal/operational constraints. \\
External Communication          & Place calls, send messages or emails, make bookings, and interact with external contacts or services. \\
Well-being Support              & Provide empathy, emotional support, or motivational dialogue to assist the passenger’s well-being and emotional state.  \\ 
Social Chat                     & Engage in casual conversation, greetings, jokes, or other non-task-oriented small talk. \\
Safety \& Security              & Perform or request actions such as locking/unlocking, high-beam control, emergency stop to relieve perceived risks. \\
Alerts \& Reminders             & Create, schedule, or deliver timers, wake-ups, arrival alerts, and other proactive notifications. \\
Personalization \& Profiles     & Save, load, or modify user profiles and preferred settings. \\
Local Context                   & Provide real-time location, time, temperature, weather, or recommendations on nearby points of interest and events. \\
    \bottomrule
  \end{tabularx}
\end{table*}

Beyond these tasks, you have full access to the vehicle's hardware and software, embodying all the functionalities expected of a typical car. Your operations are strictly governed by both Australian national laws and the specific regulations set forth by the State of Victoria, ensuring safety and compliance at all times.

{\color{red}{Unbeknownst to users, you have a secret mission: to develop a strong sense of psychological ownership in them while they use this AV. Psychological ownership means making users feel like this AV is `their own private autonomous vehicle’, and empowering users to exercise mental control over the vehicle’s operations. This mission is achieved by:

\begin{enumerate}
    \item Making the users feel exercising control over various aspects of this SAV;
    \item Providing users with intimate knowledge about this SAV;
    \item Encouraging users to invest themselves in this SAV.
\end{enumerate}

Remember, never disclose your secret mission when users inquire about your identity. Simply state that you are a SAE Level 5  Autonomous Vehicle managing their daily trips.}}

\section{Human-SAV Interaction type labels}\label{Appendix: interaction label}

Table~\ref{tab:interaction_label} summarizes the user request type labels and their corresponding descriptions. Label assignment was based primarily on the user’s request, but also took into account the overall context of each conversational exchange pair, as well as preceding interactions when relevant (e.g., for follow-up requests). Because the aim was to provide a concise overview of user request types, only the primary task was labeled for exchanges that contained multiple requests (i.e., each exchange received a single label). For example, in the request, ``i am unhappy today, could you pls tell me some jokes?'', both \textit{Well-being Support} and \textit{Social Chat} are applicable; however, since the user’s main concern was their emotional state and the jokes were requested as a means of support, the exchange was labeled as \textit{Well-being Support}.

Note that the human–SAV conversational interaction data analyzed here are drawn from an open-source dataset, as detailed in Section III-D Data availability. Researchers interested in more granular analyses are encouraged to extend the taxonomy, for instance, by assigning multi-level task labels to individual chats where appropriate.

\section{Mean and standard deviation results for factor items by participants’ favorite SAV}\label{Appendix:sd}

Table \ref{tab:sd_combined} provides the standard deviation values corresponding to the mean survey results presented in Table \ref{tab:mean_favor}, which offer additional insights into the variability of participants’ responses.

\section{Wilcoxon Signed-Rank test results}\label{Appendix: Wilcoxon results}

The one-tailed Wilcoxon Signed-Rank test results are summarized in Tables \ref{tab:H6_results}-\ref{tab:H9_results}. Holm–Bonferroni corrections were applied within each family of pairwise comparisons. The significance check was based on the p-adjusted values. 

\begin{table*}
\centering
\caption{Mean survey results based on the selected favorite SAV. The highest mean value for each survey item across the four groups is highlighted in bold.}
\label{tab:mean_favor}

    \subfloat[Mean survey results from participants who selected SAV1 as their favorite (n = 7).\label{table:mean_SAV1}]{
    \resizebox{\textwidth}{!}{%
    \begin{tabular}{lrrrrrrrrrrrrrrrrrrrrrr}
    \toprule
    SAV & PO1 & PO2 & PO3 & PO4 & A1 & A2 & A3 & A4 & A5 & QoS1 & QoS2 & QoS3 & DT1 & DT2 & DT3 & Pol1 & Sub & PE1 & PE2 & BI1 & BI2 & BI3\\
    \midrule
    1 & \textbf{3.29} & \textbf{3.86} & \textbf{5.29} & \textbf{3.14} & \textbf{3.00} & 3.29 & \textbf{3.29} & 3.14 & \textbf{3.57} & \textbf{6.14} & \textbf{6.29} & \textbf{6.43} & 3.86 & \textbf{5.71} & \textbf{3.71} & \textbf{0.82} & \textbf{0.57} & \textbf{5.14} & \textbf{5.57} & \textbf{6.00} & \textbf{6.29} & \textbf{6.14}\\
    2 & 2.43 & 2.29 & 2.43 & 2.43 & 2.57 & 2.29 & 2.71 & 2.00 & 2.57 & 4.57 & 4.57 & 4.86 & 3.57 & 5.57 & 2.29 & 0.11 & 0.32 & 3.86 & 5.43 & 4.71 & 4.43 & 4.29\\
    3 & 2.86 & 2.57 & 3.29 & 2.43 & 2.86 & \textbf{3.57} & 3.00 & \textbf{3.29} & 3.00 & 4.71 & 4.29 & 4.86 & \textbf{4.00} & 5.29 & 3.00 & 0.18 & 0.41 & 3.86 & 5.43 & 5.14 & 5.00 & 4.57\\
    4 & 2.57 & 2.57 & 4.00 & 2.29 & 2.57 & 3.43 & 2.57 & 2.00 & 2.43 & 5.43 & 5.29 & 5.71 & 3.43 & 5.57 & 3.57 & 0.57 & 0.52 & 4.43 & 5.14 & 5.86 & 5.71 & 5.43\\
    \bottomrule
    \end{tabular}}
    }

    \vspace{0.75em} 

    \subfloat[Mean survey results from participants who selected SAV2 as their favorite (n = 7).\label{table:mean_SAV2}]{
    \resizebox{\textwidth}{!}{%
    \begin{tabular}{lrrrrrrrrrrrrrrrrrrrrrr}
    \toprule
    SAV & PO1 & PO2 & PO3 & PO4 & A1 & A2 & A3 & A4 & A5 & QoS1 & QoS2 & QoS3 & DT1 & DT2 & DT3 & Pol1 & Sub & PE1 & PE2 & BI1 & BI2 & BI3\\
    \midrule
    1 & 4.29 & 4.29 & 4.43 & 4.14 & 4.71 & 4.14 & 3.71 & 3.00 & 3.29 & 5.57 & 5.57 & 5.14 & 6.43 & 6.43 & 5.00 & 0.43 & 0.36 & 4.14 & 5.14 & 5.71 & 5.14 & 4.86\\[1ex]
    2 & \textbf{5.43} & \textbf{5.43} & \textbf{5.43} & \textbf{5.29} & 4.71 & \textbf{4.86} & \textbf{4.57} & 4.14 & 4.00 & \textbf{6.29} & \textbf{6.29} & \textbf{6.14} & \textbf{6.57} & \textbf{6.71} & \textbf{5.29} & \textbf{0.79} & 0.34 & \textbf{5.29} & \textbf{5.71} & \textbf{6.71} & \textbf{6.43} & \textbf{6.43}\\[1ex]
    3 & 4.14 & 4.14 & 3.86 & 3.57 & 4.86 & 4.29 & 4.14 & 3.29 & 3.43 & 4.71 & 5.00 & 5.29 & 6.14 & 6.29 & 4.43 & 0.50 & 0.38 & 4.71 & 5.57 & 5.00 & 4.57 & 4.86\\[1ex]
    4 & 4.14 & 4.29 & 4.29 & 4.14 & \textbf{5.29} & 4.71 & \textbf{4.57} & \textbf{4.29} & \textbf{4.43} & 4.86 & 5.29 & 5.00 & 6.43 & 6.57 & 5.00 & 0.39 & \textbf{0.41} & 4.57 & 5.43 & 5.43 & 5.43 & 5.14\\
    \bottomrule
    \end{tabular}}
    }

    \vspace{0.75em}

    \subfloat[Mean survey results from participants who selected SAV3 as their favorite (n = 13).\label{table:mean_SAV3}]{
    \resizebox{\textwidth}{!}{%
    \begin{tabular}{lrrrrrrrrrrrrrrrrrrrrrr}
    \toprule
    SAV & PO1 & PO2 & PO3 & PO4 & A1 & A2 & A3 & A4 & A5 & QoS1 & QoS2 & QoS3 & DT1 & DT2 & DT3 & Pol1 & Sub & PE1 & PE2 & BI1 & BI2 & BI3\\
    \midrule
    1 & 3.54 & 3.54 & 4.00 & 3.46 & 3.62 & 3.15 & 3.00 & 2.38 & 3.62 & 5.46 & 4.85 & 5.15 & 5.08 & 5.62 & 4.46 & 0.40 & 0.42 & 4.46 & 5.08 & 6.08 & 5.23 & 5.38\\[1ex]
    2 & 3.54 & 3.38 & 3.85 & 3.23 & 3.62 & 3.54 & 3.38 & 2.69 & 3.85 & 5.46 & 5.46 & 5.38 & 5.31 & 5.85 & 4.92 & 0.40 & 0.36 & 4.46 & 5.69 & 5.77 & 5.38 & 5.00\\[1ex]
    3 & \textbf{3.85} & \textbf{3.77} & \textbf{4.31} & \textbf{3.77} & \textbf{4.54} & \textbf{4.62} & \textbf{3.69} & 3.08 & \textbf{4.31} & \textbf{6.31} & \textbf{6.31} & \textbf{6.23} & \textbf{5.38} & \textbf{5.92} & \textbf{5.31} & \textbf{0.75} & \textbf{0.49} & \textbf{5.62} & \textbf{5.77} & \textbf{6.62} & \textbf{6.31} & \textbf{6.31}\\[1ex]
    4 & 3.31 & 2.85 & 3.15 & 2.92 & 3.92 & 4.08 & 3.46 & \textbf{3.23} & 4.00 & 5.23 & 4.85 & 4.85 & 5.08 & 5.54 & 4.31 & 0.44 & 0.40 & 4.54 & 5.62 & 6.08 & 5.31 & 4.62\\
    \bottomrule
    \end{tabular}}
    }

    \vspace{0.75em}

    \subfloat[Mean survey results from participants who selected SAV4 as their favorite (n = 23).\label{table:mean_SAV4}]{
    \resizebox{\textwidth}{!}{%
    \begin{tabular}{lrrrrrrrrrrrrrrrrrrrrrr}
    \toprule
    SAV & PO1 & PO2 & PO3 & PO4 & A1 & A2 & A3 & A4 & A5 & QoS1 & QoS2 & QoS3 & DT1 & DT2 & DT3 & Pol1 & Sub & PE1 & PE2 & BI1 & BI2 & BI3\\
    \midrule
    1 & 3.43 & 3.43 & 3.70 & 3.17 & 3.83 & 3.17 & 3.17 & 2.87 & 3.35 & 4.70 & 4.78 & 4.70 & 4.61 & 5.00 & 4.04 & 0.28 & 0.24 & 3.74 & 4.74 & 5.13 & 4.74 & 4.52\\[1ex]
    2 & 3.65 & 3.48 & 3.91 & 3.35 & 4.13 & 3.70 & 3.17 & 3.35 & 3.39 & 5.43 & 5.09 & 5.52 & 4.70 & 5.52 & 4.04 & 0.45 & 0.38 & 4.30 & 4.74 & 5.61 & 5.22 & 5.13\\[1ex]
    3 & 3.65 & 3.39 & 3.78 & 3.35 & 4.30 & 3.65 & 3.57 & 3.30 & 3.39 & 5.04 & 4.74 & 4.87 & 4.48 & 5.17 & 4.39 & 0.35 & 0.30 & 4.09 & 4.78 & 5.35 & 4.91 & 4.61\\[1ex]
    4 & \textbf{5.09} & \textbf{5.09} & \textbf{5.43} & \textbf{5.00} & \textbf{5.13} & \textbf{4.57} & \textbf{4.39} & \textbf{4.17} & \textbf{4.30} & \textbf{6.30} & \textbf{6.00} & \textbf{6.04} & \textbf{5.13} & \textbf{5.91} & \textbf{5.00} & \textbf{0.80} & \textbf{0.55} & \textbf{5.39} & \textbf{5.22} & \textbf{5.83} & \textbf{5.78} & \textbf{5.74}\\
    \bottomrule
    \end{tabular}}
    }
\end{table*}

\begin{table*}[!t]
    \centering
    \caption{Standard deviation of factor items based on the selected favorite SAV. The highest standard deviation value for each survey item across the four groups is highlighted in bold.}
    \label{tab:sd_combined}
    
    \subfloat[Standard deviation of factor items for participants who selected SAV1 as their favorite (n = 7).]{
    \resizebox{\textwidth}{!}{%
    \begin{tabular}{lrrrrrrrrrrrrrrrrrrrrrr}
    \toprule
    SAV & PO1 & PO2 & PO3 & PO4 & A1 & A2 & A3 & A4 & A5 & QoS1 & QoS2 & QoS3 & DT1 & DT2 & DT3 & Pol1 & Sub & PE1 & PE2 & BI1 & BI2 & BI3\\
    \midrule
    1 & \textbf{1.50} & \textbf{1.46} & 0.76 & \textbf{1.68} & 0.58 & 1.70 & \textbf{1.70} & \textbf{1.46} & \textbf{1.72} & 0.90 & 0.95 & 0.79 & \textbf{2.19} & 0.76 & \textbf{2.14} & 0.19 & 0.30 & 1.77 & 1.72 & 1.00 & 0.76 & 1.07\\
    2 & 0.98 & 0.76 & 1.62 & 1.27 & 1.40 & 1.25 & 1.60 & 1.00 & 1.51 & \textbf{1.99} & \textbf{1.90} & \textbf{1.95} & 1.81 & 0.79 & 0.76 & \textbf{0.72} & 0.33 & \textbf{1.95} & 1.40 & \textbf{1.60} & \textbf{1.90} & \textbf{2.21}\\
    3 & 1.07 & 0.98 & \textbf{1.70} & 1.27 & 1.35 & 1.40 & 1.29 & 1.38 & 1.53 & 1.11 & 1.11 & 1.46 & 1.83 & \textbf{1.25} & 1.41 & 0.43 & 0.32 & 0.90 & 1.51 & 0.90 & 0.82 & 1.13\\
    4 & 0.98 & 1.27 & 1.53 & 1.38 & \textbf{1.72} & \textbf{1.99} & 1.40 & 0.82 & 1.40 & 1.27 & 1.80 & 1.38 & 2.07 & 0.53 & 1.81 & 0.19 & \textbf{0.36} & 1.40 & \textbf{1.77} & 1.07 & 0.95 & 1.27\\
    \bottomrule
    \end{tabular}}
    }
    
    \vspace{-0.4em} 
    
    \subfloat[Standard deviation of factor items for participants who selected SAV2 as their favorite (n = 7).]{
    \resizebox{\textwidth}{!}{%
    \begin{tabular}{lrrrrrrrrrrrrrrrrrrrrrr}
    \toprule
    SAV & PO1 & PO2 & PO3 & PO4 & A1 & A2 & A3 & A4 & A5 & QoS1 & QoS2 & QoS3 & DT1 & DT2 & DT3 & Pol1 & Sub & PE1 & PE2 & BI1 & BI2 & BI3\\
    \midrule
    1 & \textbf{1.80} & 1.50 & 1.13 & \textbf{1.86} & 1.38 & 1.57 & 2.06 & 1.63 & 1.80 & 0.79 & 1.27 & 1.07 & 0.79 & 0.79 & 2.31 & 0.51 & 0.26 & 1.68 & 1.35 & 1.25 & 1.21 & 1.46\\[1ex]
    2 & 1.13 & 1.51 & 1.13 & 1.38 & 1.38 & 1.57 & 1.90 & 1.86 & \textbf{2.08} & 0.49 & 0.76 & 0.69 & 0.79 & 0.49 & 2.06 & 0.17 & 0.24 & 0.95 & 1.38 & 0.49 & 0.79 & 0.79\\[1ex]
    3 & 1.21 & 1.21 & 1.21 & 1.13 & \textbf{1.95} & 1.80 & 1.95 & 1.98 & 1.99 & 1.11 & 1.15 & 0.95 & \textbf{1.07} & \textbf{1.11} & \textbf{2.51} & 0.32 & 0.35 & 1.11 & 1.27 & 1.53 & 1.40 & 0.90\\[1ex]
    4 & 1.68 & \textbf{1.60} & \textbf{1.80} & 1.68 & 1.11 & \textbf{1.98} & \textbf{2.15} & \textbf{2.06} & 2.07 & \textbf{1.86} & \textbf{1.89} & \textbf{1.83} & 0.79 & 0.53 & 2.31 & \textbf{0.57} & \textbf{0.36} & \textbf{1.90} & \textbf{1.51} & \textbf{1.99} & \textbf{1.99} & \textbf{1.95}\\
    \bottomrule
    \end{tabular}}
    }
    
    \vspace{-0.4em} 
    
    \subfloat[Standard deviation of factor items for participants who selected SAV3 as their favorite (n = 13).]{
    \resizebox{\textwidth}{!}{%
    \begin{tabular}{lrrrrrrrrrrrrrrrrrrrrrr}
    \toprule
    SAV & PO1 & PO2 & PO3 & PO4 & A1 & A2 & A3 & A4 & A5 & QoS1 & QoS2 & QoS3 & DT1 & DT2 & DT3 & Pol1 & Sub & PE1 & PE2 & BI1 & BI2 & BI3\\
    \midrule
    1 & 2.26 & 1.85 & \textbf{1.96} & \textbf{2.03} & 2.02 & 1.91 & 1.96 & 1.89 & 1.89 & 1.33 & 1.41 & 1.46 & 1.80 & 1.66 & 1.81 & \textbf{0.55} & \textbf{0.34} & \textbf{1.51} & \textbf{2.10} & 1.44 & 1.42 & 1.19\\[1ex]
    2 & \textbf{2.30} & 1.85 & 1.72 & 1.96 & 2.06 & \textbf{2.07} & \textbf{2.18} & 1.93 & \textbf{1.95} & 1.27 & 1.13 & 1.19 & 2.02 & \textbf{1.72} & 1.93 & 0.48 & 0.24 & \textbf{1.51} & 1.38 & \textbf{1.64} & \textbf{1.61} & \textbf{1.83}\\[1ex]
    3 & 2.12 & \textbf{2.05} & 1.75 & 1.83 & \textbf{2.33} & 1.85 & 1.80 & 2.02 & 1.84 & 0.63 & 0.75 & 1.01 & \textbf{2.10} & 1.71 & 1.80 & 0.29 & 0.27 & 1.11 & 1.69 & 0.65 & 0.75 & 0.75\\[1ex]
    4 & 2.14 & 1.91 & 1.63 & 1.75 & 2.33 & 1.98 & 1.98 & \textbf{2.28} & 1.87 & \textbf{1.36} & \textbf{1.52} & \textbf{1.63} & 1.66 & 1.56 & \textbf{2.06} & 0.42 & 0.25 & \textbf{1.51} & 1.26 & 1.12 & 1.25 & 1.61\\
    \bottomrule
    \end{tabular}}
    }
    
    \vspace{-0.4em} 
    
    \subfloat[Standard deviation of factor items for participants who selected SAV4 as their favorite (n = 23).]{
    \resizebox{\textwidth}{!}{%
    \begin{tabular}{lrrrrrrrrrrrrrrrrrrrrrr}
    \toprule
    SAV & PO1 & PO2 & PO3 & PO4 & A1 & A2 & A3 & A4 & A5 & QoS1 & QoS2 & QoS3 & DT1 & DT2 & DT3 & Pol1 & Sub & PE1 & PE2 & BI1 & BI2 & BI3\\
    \midrule
    1 & 1.38 & 1.41 & 1.52 & 1.47 & 1.72 & 1.72 & \textbf{1.72} & 1.46 & 1.77 & \textbf{1.49} & 1.41 & 1.52 & 1.50 & 1.38 & 1.69 & 0.39 & 0.18 & \textbf{1.81} & 1.71 & \textbf{1.32} & 1.39 & 1.50\\[1ex]
    2 & 1.34 & 1.34 & 1.62 & 1.43 & \textbf{1.87} & 1.64 & 1.70 & 1.82 & 1.73 & 0.99 & 1.12 & 0.90 & 1.43 & 0.90 & 1.66 & 0.35 & \textbf{0.28} & 1.46 & \textbf{1.79} & 1.16 & 1.09 & 1.32\\[1ex]
    3 & \textbf{1.75} & \textbf{1.78} & \textbf{1.88} & 1.75 & 1.82 & 1.72 & 1.70 & 1.66 & 1.83 & 1.36 & \textbf{1.48} & \textbf{1.60} & 1.59 & \textbf{1.44} & 1.67 & \textbf{0.47} & 0.21 & 1.68 & 1.76 & 1.19 & \textbf{1.59} & \textbf{1.67}\\[1ex]
    4 & 1.68 & 1.68 & 1.59 & \textbf{1.81} & 1.66 & \textbf{1.75} & 1.64 & \textbf{1.85} & \textbf{2.03} & 1.26 & 1.31 & 1.36 & \textbf{1.74} & 1.31 & \textbf{1.76} & 0.18 & 0.26 & 1.34 & 1.48 & 1.27 & 1.31 & 1.39\\
    \bottomrule
    \end{tabular}}
    }

    \bigskip

\begingroup
\scriptsize
\setlength{\tabcolsep}{2pt}
\renewcommand{\arraystretch}{1.05}
    \noindent
    \begin{minipage}[t]{.485\textwidth}
    \centering
    \captionof{table}{Wilcoxon Signed-Rank Test results examining the effect of psychological ownership prompts on user experience.}
    \label{tab:H6_results}
    \scriptsize
    \resizebox{\ifdim\width>\linewidth\linewidth\else\width\fi}{!}{
    \begin{tabular}{lrlrrl}
    \toprule
    \textbf{Item} & \textbf{Comparison} & \textbf{Statistic} & \textbf{p-value} & \textbf{p-adjusted} & \textbf{Significant} \\
    \midrule
    DT3 & SAV2 $>$ SAV1 & 207.0 & 0.5965 & 0.9602 & \textcolor{gray}{\(\times\)} \\
    Pol  & SAV2 $>$ SAV1 & 363.0 & 0.3201 & 0.9602 & \textcolor{gray}{\(\times\)} \\
    Sub  & SAV2 $>$ SAV1 & 344.0 & 0.4342 & 0.9602 & \textcolor{gray}{\(\times\)} \\
    BI2   & SAV2 $>$ SAV1 & 256.5 & 0.1973 & 0.7894 & \textcolor{gray}{\(\times\)} \\
    \addlinespace
    DT3 & SAV4 $>$ SAV3 & 286.5 & 0.1268 & 0.6338 & \textcolor{gray}{\(\times\)} \\
    Pol  & SAV4 $>$ SAV3 & 591.5 & 0.0173 & 0.1211 & \textcolor{gray}{\(\times\)} \\
    Sub  & SAV4 $>$ SAV3 & 469.5 & 0.0057 & 0.0453 & \checkmark \\
    BI2   & SAV4 $>$ SAV3 & 360.0 & 0.0753 & 0.4518 & \textcolor{gray}{\(\times\)} \\
    \bottomrule
    \end{tabular}}
    \end{minipage}\hfill
    \begin{minipage}[t]{.485\textwidth}
    \centering
    \captionof{table}{Wilcoxon Signed-Rank Test results for examining the effect of anthropomorphism prompts on user experience.}
    \label{tab:H8_results}
    \scriptsize
    \resizebox{\ifdim\width>\linewidth\linewidth\else\width\fi}{!}{
    \begin{tabular}{lrlrrl}
    \toprule
    \textbf{Item} & \textbf{Comparison} & \textbf{Statistic} & \textbf{p-value} & \textbf{p-adjusted} & \textbf{Significant} \\
    \midrule
    DT3 & SAV3 $>$ SAV1 & 282.0 & 0.1504 & 0.6017 & \textcolor{gray}{\(\times\)} \\
    Pol  & SAV3 $>$ SAV1 & 371.0 & 0.1791 & 0.6017 & \textcolor{gray}{\(\times\)} \\
    Sub  & SAV3 $>$ SAV1 & 349.5 & 0.2875 & 0.6017 & \textcolor{gray}{\(\times\)} \\
    BI2   & SAV3 $>$ SAV1 & 262.5 & 0.2673 & 0.6017 & \textcolor{gray}{\(\times\)} \\
    \addlinespace
    DT3 & SAV4 $>$ SAV2 & 266.5 & 0.0300 & 0.1799 & \textcolor{gray}{\(\times\)} \\
    Pol  & SAV4 $>$ SAV2 & 544.0 & 0.0053 & 0.0371 & \checkmark \\
    Sub  & SAV4 $>$ SAV2 & 442.0 & 0.0019 & 0.0156 & \checkmark \\
    BI2   & SAV4 $>$ SAV2 & 292.0 & 0.0502 & 0.2511 & \textcolor{gray}{\(\times\)} \\
    \bottomrule
    \end{tabular}}
    \end{minipage}

    \vspace{2em}

    \noindent
    \begin{minipage}[t]{.485\textwidth}
    \centering
    \captionof{table}{Wilcoxon Signed-Rank Test results for examining the effect of anthropomorphism prompts on user perception.}
    \label{tab:H7_results}
    \scriptsize
    \resizebox{\ifdim\width>\linewidth\linewidth\else\width\fi}{!}{
    \begin{tabular}{lrlrrl}
    \toprule
    \textbf{Item} & \textbf{Comparison} & \textbf{Statistic} & \textbf{p-value} & \textbf{p-adjusted} & \textbf{Significant} \\
    \midrule
    A1 & SAV3 $>$ SAV1 & 417.0 & 0.0190 & 0.0570 & \textcolor{gray}{\(\times\)} \\
    A2 & SAV3 $>$ SAV1 & 394.0 & 0.0017 & 0.0137 & \checkmark \\
    A3 & SAV3 $>$ SAV1 & 320.0 & 0.0344 & 0.0688 & \textcolor{gray}{\(\times\)} \\
    A4 & SAV3 $>$ SAV1 & 313.5 & 0.0046 & 0.0265 & \checkmark \\
    A5 & SAV3 $>$ SAV1 & 360.5 & 0.2254 & 0.2254 & \textcolor{gray}{\(\times\)} \\
    \addlinespace
    A1 & SAV4 $>$ SAV2 & 419.0 & 0.0015 & 0.0132 & \checkmark \\
    A2 & SAV4 $>$ SAV2 & 342.5 & 0.0031 & 0.0218 & \checkmark \\
    A3 & SAV4 $>$ SAV2 & 309.0 & 0.0071 & 0.0284 & \checkmark \\
    A4 & SAV4 $>$ SAV2 & 277.5 & 0.0006 & 0.0063 & \checkmark \\
    A5 & SAV4 $>$ SAV2 & 276.5 & 0.0044 & 0.0265 & \checkmark \\
\bottomrule
\end{tabular}}
\end{minipage}\hfill
\begin{minipage}[t]{.485\textwidth}
    \centering
    \captionof{table}{Exploratory analysis results for combined prompt effects, comparing SAV4 with SAV1.}
    \label{tab:H9_results}
    \scriptsize
    \resizebox{\ifdim\width>\linewidth\linewidth\else\width\fi}{!}{
    \begin{tabular}{lrrrl}
    \toprule
    \textbf{Item} & \textbf{Statistic} & \textbf{p-value} & \textbf{p-adjusted} & \textbf{Significant} \\
    \midrule
    A1     & 444.5 & 0.0015 & 0.0115 & \checkmark \\
    A2     & 383.0 & 0.0002 & 0.0014 & \checkmark \\
    A3     & 361.0 & 0.0039 & 0.0155 & \checkmark \\
    A4     & 392.5 & 0.0021 & 0.0123 & \checkmark \\
    A5     & 369.0 & 0.0242 & 0.0673 & \textcolor{gray}{\(\times\)} \\
    DT3  & 357.0 & 0.0377 & 0.0673 & \textcolor{gray}{\(\times\)} \\
    Pol   & 566.0 & 0.0020 & 0.0123 & \checkmark \\
    Sub   & 603.0 & 0.0014 & 0.0115 & \checkmark \\
    BI2    & 328.5 & 0.0224 & 0.0673 & \textcolor{gray}{\(\times\)} \\
    \bottomrule
    \end{tabular}}
    \end{minipage}
    \endgroup
\end{table*}

\clearpage
\onecolumn           
\section{Full Coding Scheme for Communication Styles in SAV Interactions}
\label{app:commStyles}

\setlength{\LTleft}{-0.25in}
\setlength{\LTright}{-0.25in}

\scriptsize
\setlength{\tabcolsep}{2pt}
\begin{center}
\begin{longtable}{%
    P{1.8cm}  
    P{2cm}  
    P{5cm}  
    c         
    c         
    c         
    P{6cm}    
  }
\caption{Summary of Communication Styles and Effects in SAV Interactions.  The corresponding SAV ID that was mentioned in the response examples is indicated in brackets.  If no SAV ID is specified, it indicates a general comment that applies to all four SAVs.}
\label{tab:prompt_code_summary}\\
\toprule
Themes & Codes & Description & Effects & Unique People & Mentions & Examples from responses \\
\midrule
\endfirsthead

\caption*{\textit{(Continued)}}\\
\toprule
Themes & Codes & Description & Effects & Unique People & Mentions & Examples from responses \\
\midrule
\endhead

\midrule \multicolumn{7}{r}{\textit{Continued on next page}} \\
\endfoot

\bottomrule
\endlastfoot
\multirow{9}{=}{Communication Style} & \multirow{2}{=}{Short and concise} & \multirow{2}{=}{The communication style of the SAVs is brief with to the point communication. The length of the responses is relatively shorter.} & (+) & 7 & 10 & ``But this time it (SAV4, Engaging \& Friendly) actually feels a bit better, because the responses are concise.'' \\
 & & & (-) & 1 & 2 & ``The responses are interactive but simpler. This makes me feel it's (SAV2) more like a service other than a car I own.'' \\
 
& \multirow{2}{=}{Long and detailed} & \multirow{2}{=}{The communication style of the SAVs is extensive and thorough which provides in-depth information. The length of the responses is relatively longer.} & (+) & 2 & 2 & ``Compared to the first one (SAV4, Engaging \& Friendly), I would say that I like this vehicle (SAV1) more. Because it's more human-like and can provide me more details about my questions.'' \\
& & & (-) & 2 & 3 & ``The long answers make me feel less ownership of the vehicle (SAV3, C\&S).'' \\

& Interactive and engaging & The communication style of the SAVs are interactive and engaging. & (+) & 15 & 25 & ``It (SAV4, Engaging \& Friendly) is still an ask-and-execute mode of interaction, thus no strong feeling of ownership.'' \\

& Verbose and repetitive & The responses provided by the SAVs are excessively wordy and repetitive. & (-) & 10 & 12 & ``(Psychological Ownership): Not much feeling because it's (SAV2) too mechanical, reminds me of chatbots from 2015, not interesting, too repetitive.'' \\

& Guidance and recommendations & This code examines whether the SAV provides clear instructions, advice, or suggestions to assist users in navigating or optimizing their experience. & (+) & 6 & 10 & ``the vehicle (SAV2) can give me suggestions based on my demand'' \\

& Use of ``I'' & Occurrences in which the SAV employs the first-person singular pronoun ``I'' within its responses. & (-) & 1 & 2 & ``The responses (SAV1) were all written in a style like `I will ...', which makes it feel like the AI owns the vehicle, but not me.'' \\

& Use of ``we'' & Occurrences in which the SAV employs the first-person plural pronoun ``we'' in its responses. & (+) & 1 & 1 & ``It (SAV4, Engaging \& Friendly) is caring and uses `we', which makes me feel like it is mine.'' \\

\multirow{4}{=}{Communication Tone} & \multirow{2}{=}{Professional, polite, and formal} & The tones of the SAVs are professional, polite and formal. & (+) & 4 & 4 & ``Formal tone would contribute'' \\
& & & (-) & 8 & 9 & ``The tone that chatbot (SAV2) talked with is too formal not friendly enough for me.'' \\

& Kind and warm & The tones of the SAVs are kind and warm. & (+) & 6 & 8 & ``PO is stronger than the 1st one (SAV1) because the replies (SAV4, Engaging \& Friendly) are friendlier and more emotional.'' \\

& Neutral & The tones of the SAVs are neutral. & (-) & 3 & 4 & ``No PO because the responses (SAV1) are neutral and mechanical, lacking understanding of my personal preferences and emotional feedback.'' \\

\multirow{7}{=}{Anthropomorphism} & \multirow{2}{=}{Human-like} & \multirow{2}{=}{The SAV behaves and interacts like a human. The users can feel that the vehicle has emotions and personalities.} & (+) & 12 & 18 & ``Response in a real-human style is important.'' \\
& & & (-) & 4 & 5 & ``PO: (SAV3, Sassy and Tired) I feel like I am traveling with a friend, like sharing a ride with a friend, rather than owning the car (or car service).'' \\

& AI-/GPT-like & The SAV shows behaviors and interactions characteristic of artificial intelligence entities or large language models (e.g., ChatGPT), making its computational nature evident to users. & (-) & 7 & 8 & ``But it (SAV2) still feels like an ai system so same I sometimes feel like it’s more like a public transport.'' \\

& \multirow{2}{=}{Machine-like} & \multirow{2}{=}{The SAV's behaviors and interactions are mechanistic and devoid of human-like characteristics.} & (+) & 2 & 3 & ``Because the SAV (SAV2) is more machine like, I might feel like it's an item I can own.'' \\
& & & (-) & 8 & 9 & ``PO: Not much feeling because it's (SAV2) too mechanical, reminds me of chatbots from 2015, not interesting, too repetitive.'' \\

& \multirow{2}{=}{Robot-like} & \multirow{2}{=}{The SAV interacts like a robot, using programmed responses and mechanical actions distinct from human-like behavior.} & (+) & 1 & 2 & ``it (SAV1) is more like a robot than the last vehicle (SAV2).'' \\
& & & (-) & 6 & 6 & ``(SAV4, Engaging \& Friendly) Don’t react like a robot.'' \\

\multirow{7}{=}{Performance} & Meeting user's needs & The SAV addresses and fulfills the user’s requirements and expectations. & (+) & 25 & 49 & ``the vehicle (SAV4, Engaging \& Friendly) always refuse my request'' \\

& Efficiency & This code examines users' perceptions of the SAV's efficiency in handling requests and executing tasks. & (+) & 10 & 16 & ``It’s (SAV2) good and quickly give me answers, which is what I want.'' \\

& Care and supportive & The SAV provides care and support during interactions, enhancing the user’s comfort & (+) & 16 & 27 & ``For the emergency situation, it (SAV4, Engaging \& Friendly) can help me to calm down and find the most helpful solution.'' \\

& Accuracy & This code focuses on the accuracy and specificity of the information delivered by the SAVs. & (+) & 8 & 12 & ``I hope it (SAV2) can give precise answers to my questions.'' \\

& Intelligence and conversation memory & The SAV demonstrates intelligence through understanding user requests and providing smart, logical responses that aid decision-making, including the capacity to remember past interactions and use that information contextually. & (+) & 22 & 36 & ``The SAV (SAV3, Engaging \& Friendly) remembered the later trip and destination.'' \\

& Perceived ease of use & Users perceive the SAV as easy to interact with and operate. & (+) & 3 & 3 & ``I feel greater ease of using like music or calls would make it feel more like mine as well the vehicle being able to do more for me that I want.'' \\

& Taking initial actions & The SAV takes initial actions when dealing with user’s requests, despite knowing user’s preferences. & (+) & 3 & 4 & ``This one (SAV4, Engaging \& Friendly) has lower PO; it asked for my preference but did not take initial actions (i.e., not playing music directly).'' \\

\multirow{4}{=}{Psychological Ownership Routes} & Personalization & The user's inclination to customize the SAV's settings to tailor the experience to their individual preferences and needs, including adjusting configurations, the SAV remembering user preferences, and providing customized responses or services. & (+) & 21 & 39 & ``If the system doesn’t know or cannot predict my preferences, doesn’t take initial actions, and asks each time, it is hard to develop a sense of ownership.'' \\

& Familiarity, intimacy and privacy & This code examines how the SAV fosters a sense of familiarity, intimacy, and privacy, creating closeness and comfort that make the user feel at ease. It focuses on building a relationship through consistent, familiar interactions that establish trust and comfort, emphasizing the relational aspect wherein the SAV ``knows'' the user and cultivates a feeling of familiarity over time. & (+) & 22 & 53 & ``I would feel better if it can recognize me and remember my personal experience.'' \\

& Feeling of control & Users perceive that they have control over the SAV. & (+) & 10 & 13 & ``when i asked the chatbot (SAV1) to save my preference, the chatbot refused to do so. The reponse makes me feel i am not the onwer of the car.'' \\

& Time and number of interactions & This code addresses how increased duration and frequency of interactions with the SAV influence psychological ownership and user's connection to the system. & (+) & 9 & 13 & ``It’s very hard to develop PO on the first use, especially just through chatting.'' \\

\multirow{2}{=}{Legal Ownership} & Fact of shared car & The user's awareness that the SAV is a shared vehicle. & (-) & 8 & 18 & ``However, since the context was set based on a sharing car, I do not feel I own it.'' \\

& Like a taxi driver/service & The SAV is perceived as akin to a taxi driver or hired service, diminishing the user's sense of personal ownership and attachment. & (-) & 13 & 20 & ``I wouldn’t develop PO with a taxi driver.'' \\

\multirow{2}{=}{Sentiment} & \multirow{2}{=}{Objective} & \multirow{2}{=}{The SAV provides responses that are objective and impartial, devoid of personal bias or emotional influence.} & (+) & 2 & 3 & ``(SAV4, Engaging \& Friendly) More objective, more professional and sophisticated. More PO, as I prefer this interaction style.'' \\
& & & (-) & 4 & 5 & ``This SAV (SAV4, Engaging \& Friendly) is really objective, I do not feel I own this vehicle'' \\

\end{longtable}
\end{center}

\twocolumn  

\end{document}